\documentclass[twocolumn,english,aps,prb,twocolum,superscriptaddress,natbib,bibnotes,amsmath,amssymb,floatfix,groupedaddress,footinbib]{revtex4-2}
\usepackage[colorlinks=true,citecolor=blue,linkcolor=magenta]{hyperref}

\usepackage[addedmarkup=colored, authormarkupposition=left]{changes} 
\usepackage{soul}
\definechangesauthor[name={TJK}, color=red]{TJK}
\usepackage[english]{babel}
\usepackage{amsmath,amsfonts,amssymb}
\usepackage[T1]{fontenc}
\usepackage{xurl}

\usepackage{amsmath}
\usepackage{siunitx}
\usepackage[version=4]{mhchem}
\usepackage{amsfonts}
\usepackage{amssymb}

\usepackage{epstopdf}
\usepackage{graphicx}
\graphicspath{{./Figures/}}
\usepackage[utf8]{inputenc}
\UseRawInputEncoding

\begin{document}
	
	\title{Copper-impurity-free photonic integrated circuits enable deterministic soliton microcombs}
%	\title{Deterministic soliton microcombs enabled by copper impurity-free photonic integrated circuits}
%	Deterministic soliton microcombs enabled by copper-free photonic integrated circuits
%	\title{Deterministic soliton microcombs enabled by copper impurity-free photonic integrated circuits}

	\author{Xinru Ji$^{1,2}$, 
		Xurong Li$^{1,2}$,
		Zheru Qiu$^{1,2}$,
		Rui Ning Wang$^{3}$,
		Marta Divall$^{1,2}$,
		Andrey Gelash$^{1,2}$,
		Grigory Lihachev$^{1,2}$,
		and Tobias J. Kippenberg$^{1,2\dag}$}
	\affiliation{
		$^1$Institute of Physics, Swiss Federal Institute of Technology Lausanne (EPFL), CH-1015 Lausanne, Switzerland\\
		$^2$Institute of Electrical and Micro Engineering, Swiss Federal Institute of Technology Lausanne (EPFL), CH-1015 Lausanne, Switzerland\\
		$^3$Luxtelligence SA, CH-1015 Lausanne, Switzerland
	}
	
	\maketitle
	\noindent\textbf{
    Chip-scale optical frequency combs based on microresonators (microcombs) have provided access to optical combs with repetition rates ranging from GHz to THz, broad bandwidth, compact form factors, and compatibility with wafer-scale manufacturing \cite{kippenberg_dissipative_2018}. Si$_3$N$_4$ photonic integrated circuits emerged as a leading platform, offering low loss, a wide transparency window, lithographic dispersion engineering, and availability in commercial 200-mm-wafer foundries. They have been employed in nearly all system-level demonstrations to date, spanning from optical communications \cite{marin-palomo_microresonator-based_2017}, parallel LiDAR \cite{riemensberger_massively_2020}, optical frequency synthesis \cite{spencer_optical-frequency_2018}, low-noise microwave generation \cite{zhao_all-optical_2024}, to parallel convolutional processing \cite{feldmann_parallel_2021}.
    Yet, transitioning to real-world deployment outside laboratories has been compounded by the difficulty of deterministic soliton microcomb generation, primarily due to significant thermal instabilities.
    A variety of techniques have been developed to mitigate thermal effects, including pulsed pumping, fast scanning, and auxiliary laser pumping
    \cite{obrzud_microphotonic_2019,brasch_bringing_2016,zhang_sub-milliwatt-level_2019,zhou_soliton_2019,shen2020integrated}.
However these techniques do not eliminate thermal effects and often compromise microcomb performance, either by adding additional complexity or reducing the accessible soliton existence range.
%    An outstanding challenge has been to remove thermal effects in Si$_3$N$_4$ that compound soliton generation. 
Here we overcome thermal effects and demonstrate deterministic soliton generation in Si$_3$N$_4$ photonic integrated circuits. 
We identify the origin of thermal effects due to the surprising presence of copper impurities within the waveguides.
    These impurities originate from innate residual contaminants in CMOS-grade Si wafers, which are gettered into Si$_3$N$_4$ during fabrication.
    By developing copper removal techniques, we dramatically reduce thermal effects.
    We demonstrate successful DKS generation with arbitrary laser scanning profiles and slow laser scanning. Our techniques can be readily applied to front-end-of-line processing of Si$_3$N$_4$ devices in foundries, removing a key obstacle to the deployment of soliton microcomb technology.
	}
%The pioneering work of Kao \cite{kao1966dielectric}, which led to the development of low-loss optical fibers, underpins our information society and has given rise to optical communication networks that today span billions of kilometers of fibers.
%The pioneering work by Kao et al. \cite{kao1966dielectric} established that impurity-free silica glass could achieve ultra-low optical attenuation.
%Early low-loss optical fibers significantly reduced attenuation by removing metallic impurities (e.g., Cu, Fe) and other contaminants to sub-ppb levels, suppressing absorption across visible and infra-red wavelengths \cite{miya1979ultimate,ohishi1981impurity,mitachi_reduction_1984}. This enabled the production of high-purity fibers with sub-dB/km performance \cite{nagayama2002ultra,tamura2017lowest}, now forming the backbone of global communication networks spanning billions of kilometers of fiber.

The pioneering work by Kao et al. \cite{kao1966dielectric} demonstrated that ultra-low optical attenuation could be achieved in impurity-free silica glass. Subsequent advances in fiber technology reduced attenuation to sub-dB/km levels by eliminating metallic impurities (e.g., Cu, Fe) and other contaminants to sub-ppb (parts per billion) concentrations \cite{miya1979ultimate,ohishi1981impurity,mitachi_reduction_1984}, suppressing absorption across visible and infrared wavelengths. These breakthroughs enabled the production of high-purity fibers \cite{nagayama2002ultra,tamura2017lowest} that now form the backbone of global communication networks, with millions of kilometers deployed worldwide.
%Such ultra-purified precursors (e.g., SiCl$_4$) and refined fabrication \cite{mitachi_reduction_1984}, enabled sub-dB/km losses \cite{nagayama2002ultra}, laying the foundation for global optical networks.

%%%%%%%%%%%%%%%%%%%
In a similar vein, advances in reducing optical losses in integrated photonics, traditionally at the dB/cm level, have spurred interdisciplinary applications~\cite{thomson2016roadmap,sun2023applications,lu2024emerging}.
%enabled new operating principles~\cite{thomson2016roadmap,sun2023applications,lu2024emerging}.
Among these, ultra-low-loss silicon nitride (Si$_3$N$_4$) photonic integrated circuits (PICs), with high Kerr nonlinearity and a wide transparency window, have featured dB/m-level losses \cite{ye2019low,ji2021methods,liu_high-yield_2021,ji_efficient_2024} and enabled groundbreaking advances, particularly through: high-gain continuous-travelling-wave parametric amplifiers \cite{riemensberger_photonic_2022}, integrated analogues of erbium-doped fiber amplifiers and lasers \cite{liu_photonic_2022,liu_fully_2024}, frequency-agile lasers with phase noise on par with fiber lasers and unprecedented chirp rates \cite{li2021reaching,lihachev2022low}, and dissipative Kerr soliton (DKS) - chip-scale optical frequency combs with GHz-to-THz mode spacing.
Notably, DKS has proven particularly versatile, enabling system-level demonstrations in optical communications \cite{marin-palomo_microresonator-based_2017}, parallel LiDAR \cite{riemensberger_massively_2020}, optical frequency synthesis \cite{spencer_optical-frequency_2018}, low-noise microwave generation \cite{liu2020photonic,zhao_all-optical_2024} and parallel convolutional processing \cite{feldmann_parallel_2021}.

	\begin{figure*}[t!]
		\centering
		\includegraphics[width=\textwidth]{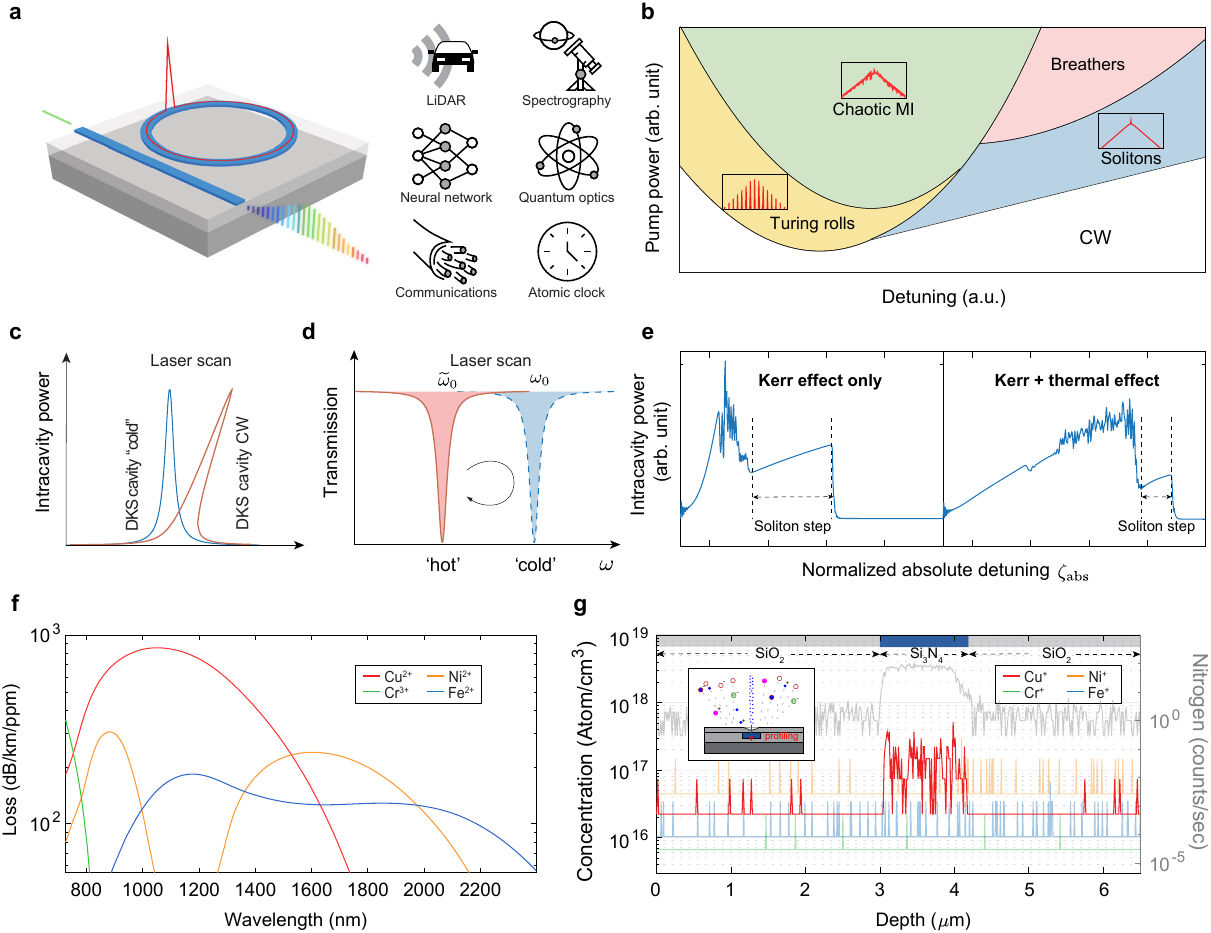}
		\caption{ 
			\footnotesize
			\textbf{Thermal absorption-induced instabilities in Dissipative Kerr soliton (DKS) generation in Si$_3$N$_4$ microresonators.} 
			\textbf{a,} Schematic of DKS generation and its diverse applications, including parallel LiDAR, astronomical spectrograph calibration, optical computing, optical communications, frequency calibration, and quantum optics.
			\textbf{b,} Stability chart of soliton microcomb states in Si$_3$N$_4$ microresonators with anomalous group velocity dispersion, as a function of laser detuning and pump power, with time and frequency domain waveforms predicted by the Lugiato-Lefever Equation (LLE).
			MI: modulation instability. CW: continuous wave. 
			\textbf{c,} Resonance tilt during pump laser scanning from blue- to red-detuned regimes. Kerr effect induces self-phase modulation, causing a tilted resonance.
			\textbf{d,} Schematic of cavity resonance shifts.
			The cold cavity resonance $\omega_0$ thermally shifts to $\widetilde{\omega}_0$ with intracavity energy (e.g., a soliton).
			The effective detuning of the laser to this shifted resonance determines the soliton properties.
			\textbf{e,} Numerical simulations of DKS generation in Si$_3$N$_4$ microresonators, illustrating the destabilizing impact of thermal effects on soliton generation.
			\textbf{f,} Transmission loss spectrum of fluoride optical fibers (in dB/km/ppm atomic), showing contributions from various metal impurities, with Cu ions exhibiting the highest absorption near 1550 nm~\cite{mitachi1983reduction}.
%			The graph is converted to dB/km/ppm (atomic).
			\textbf{g,} Depth profile of metal impurities in a fully SiO$_2$-cladded Si$_3$N$_4$ microresonator, measured by secondary ion mass spectroscopy (SIMS).
			Nitrogen is shown in grey for layer identification.
			The inset provides a graphical illustration of the SIMS measurement.
			The oxidation state in the original Si$_3$N$_4$ waveguide may differ due to the ionization process during sputtering.}
			\label{Fig:1}
	\end{figure*}

%Yet, despite these advances, little is known about the prevailing presence of thermal absorption in such ultra-low-loss Si$_3$N$_4$ waveguides.
%Despite these advances, thermal absorption in ultra-low-loss Si$_3$N$_4$ waveguides remains poorly understood.
%In fact, despite progress in achieving ultra-low loss, residual thermal absorption remains the key obstacle to deterministic access to dissipative Kerr soliton (DKS) microcombs.
Despite these advances, deterministic DKS microcomb generation remains challenging, mainly due to residual thermal effects in low-loss Si$_3$N$_4$ waveguides.
Although DKS is inherently stable once formed, it requires initialization by tuning the pump laser into the red-detuned regime of the cavity resonance \cite{herr_temporal_2014}.
This process is complicated by thermal and nonlinear dynamics \cite{carmon_dynamical_2004}, where the transition from modulation instability (MI) combs to solitons results in an intracavity power drop, inducing thermo-optic resonance shifts that narrow the soliton existence range (Fig. \ref{Fig:1}c-e, Supplementary Note 1).
%and often causing the system to revert to continuous-wave operation
Techniques such as rapid laser tuning \cite{brasch_bringing_2016}, active feedback \cite{yi_active_2016}, pulse pumping \cite{obrzud_microphotonic_2019}, auxiliary-laser pumping \cite{zhang_sub-milliwatt-level_2019, zhou_soliton_2019},
pump laser self-injection locking \cite{shen2020integrated}, pump laser modulation \cite{wildi_thermally_2019}, and dual-mode pumping \cite{li_stably_2017,weng_dual-mode_2022} have been developed to mitigate thermal effects.
While these methods facilitate single-soliton generation, they do not fundamentally eliminate thermal effects and introduce additional complexities, necessitating specialized resonator designs and constraining the effective pump detuning range.
Recently, weakly confined Si$_3$N$_4$ PICs have been employed to generate interlocked switching waves, but equally exhibit thermal effects and require fast laser scanning to initiate soliton formation \cite{ji2024dispersive,ji_multimodality_2024}.
Consequently, achieving deterministic and robust soliton access remains an unresolved challenge, limiting the scalability and practical deployment of soliton microcombs in real-world applications.

%%%%%%%%
Over the past decade, significant progress has been made in reducing optical losses in Si$_3$N$_4$ PICs \cite{liu_high-yield_2021}. Thermal absorption in Si$_3$N$_4$ PICs, has been reduced primarily through high-temperature annealing to eliminate hydrogen-related bonds (N-H, Si-H, O-H) \cite{ye2019low,ji2021methods,ji_efficient_2024}. However, the origins of residual thermal losses remain poorly understood.
Even though prior work has detected the presence of copper (Cu) impurities in Si$_3$N$_4$ waveguides \cite{pfeiffer_ultra-smooth_2018}, the origin, impact, and removal of Cu impurities in Si$_3$N$_4$ PICs remain unexplored.

Here, we identify Cu impurities as the primary source of thermal absorption in Si$_3$N$_4$ PICs, trace their origin, and devise methods to remove them.
%Using secondary ion mass spectroscopy (SIMS), we detect transition metals in Si$_3$N$_4$ waveguides; however, only Cu is present at levels sufficient to cause measurable absorption in the near-infrared range (Fig. \ref{Fig:1}f, \cite{mitachi1983reduction}).
Surprisingly, Cu contamination occurs during fabrication, even when using high-purity electronic-grade Si wafers (EG-Si, used in CMOS industry) with non-detectable surface Cu levels.
These impurities stem from Cu residing in the bulk Si wafers, which diffuse into the Si$_3$N$_4$ layer and are trapped during high-temperature annealing.
By eliminating these impurities, we realize Cu-free PICs that enable deterministic soliton generation in Si$_3$N$_4$ microresonators without complex (fast) laser tuning.
Beyond resolving a major longstanding challenge in microcomb technology, our findings pave the way for achieving even lower losses in integrated photonics.

%%%%%%%%%%%%%%%%
\subsection*{Copper Impurities: Impact on Si$_3$N$_4$ PICs and Soliton Generation}	

\begin{figure*}[t!]
		\centering
		\includegraphics[width=\textwidth]{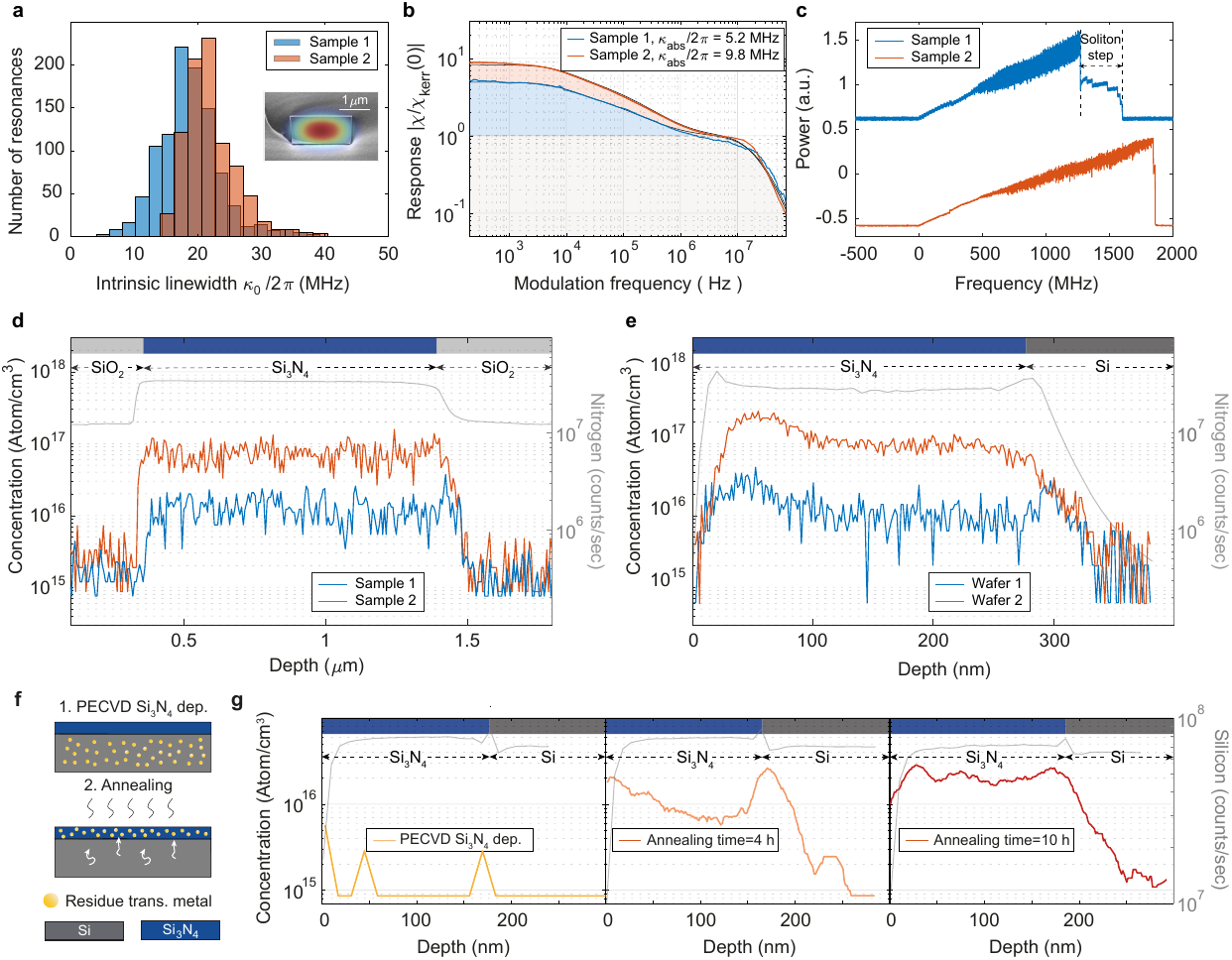}
		\caption{ 
			\footnotesize
			\textbf{Copper-induced thermal absorption loss in Si$_3$N$_4$ microresonators: origins and impacts on DKS generation.} 
			\textbf{a,} Histogram of intrinsic linewidth $\kappa_0/2\pi$ (1260-1630 nm, TE polarization), showing optical loss variations between two concurrently fabricated samples using Si wafers from different suppliers.
			Inset: Simulated optical TE$_{00}$ fundamental mode.
			Device IDs: Sample 1: \texttt{D163\_11\_F1\_C7}; Sample 2: \texttt{D163\_13\_F8\_C7}.
			\textbf{b,} Kerr-normalized thermal response measurements at resonance around 1550~nm for Sample 1 and Sample 2.
			Colored regions represent thermal contributions, which respond at lower modulation frequencies.
			\textbf{c,} Measured soliton steps during pump laser scanning in TE polarization: Sample 1 exhibits a longer soliton step ($\sim$ 330 MHz) compared to Sample 2 ($\sim$ 8.6 MHz).
			\textbf{d,} Depth-resolved Cu concentration profiles for Sample 1 and Sample 2, measured via SIMS by monitoring the Cu$^\mathrm{+}$ peak (m/z = 62.95) during ion beam sputtering.
			The SiO$_2$ cladding was partially removed to focus on the Si$_3$N$_4$ waveguides.
			SIMS results are corrected based on the Si$_3$N$_4$ filling ratio (21.76$\%$) within the mixed Si$_3$N$_4$/SiO$_2$ waveguide layer.
			\textbf{e,} Cu profiles in Si substrates used for fabricating Sample 1 (Wafer 1: Siegert Wafer) and Sample 2 (Wafer 2: Silicon Valley Microelectronics) reveal contamination levels comparable to those in the Si$_3$N$_4$ waveguides, identifying the Si substrate as the source of Cu impurities.
			The measurement was done by depositing 250-nm LPCVD Si$_3$N$_4$ on Si wafers, followed by 11-hour annealing at 1200$^{\circ}$C to concentrate diluted Cu from Si into the gettering Si$_3$N$_4$ layer.
			\textbf{f,} Schematic of the copper detection method in silicon substrates, involving plasma-enhanced chemical vapor deposition (PECVD) of 200-nm Si$_3$N$_4$ films, and high-temperature annealing (800$^{\circ}$C, 10 hours) to concentrate Cu ions for detection above the sensitivity limit.
			\textbf{g,} Cu diffusion profiles during annealing (as-deposited, 4 hours, and 10 hours), showing progressive migration from the silicon substrate to the Si$_3$N$_4$ layers.
			}
		\label{Fig:2}
	\end{figure*}

Cu, along with other transition metals, can introduce significant optical loss, which is well-studied in optical fibers (Fig. \ref{Fig:1}f) \cite{mitachi1983reduction}. Driven by the prior observation of Cu impurities in Si$_3$N$_4$ PICs \cite{pfeiffer_ultra-smooth_2018}, we use secondary ion mass spectroscopy (SIMS) to systematically investigate the impact of metal impurities on Si$_3$N$_4$ PICs with ppba sensitivity and depth resolution.
Among 14 metallic impurities, Cu is the dominant contamination in Si$_3$N$_4$ waveguides (Fig. \ref{Fig:1}g and Supplementary Note 2).
Next, the impact of Cu impurities was studied on soliton microcomb generation by two representative Si$_3$N$_4$ microresonators (Fig. \ref{Fig:2}a-d).
The fully SiO$_2$-cladded Si$_3$N$_4$ microresonators were fabricated concurrently with the photonic Damascene process~\cite{liu_high-yield_2021}, with a $2.2\times0.9$ $\mu$m$^2$ cross-section and a 100-GHz free spectral range (FSR).
The most probable intrinsic linewidths in the histogram (Fig. \ref{Fig:2}a) indicate mean losses of $18.74\pm0.17$ MHz ($\alpha = 3.57\pm0.03$ dB/m) for Sample 1 and $21.99\pm0.13$ MHz ($\alpha = 4.21\pm0.02$ dB/m) for Sample 2.
%Figure~\ref{Fig:2}a shows histograms of intrinsic linewidths $\kappa_0/2\pi$ (1260-1630 nm, TE polarization), with mean losses of 18.74$\pm$0.17 MHz ($\alpha = 3.57\pm0.03$ dB/m) for Sample 1 and 21.99$\pm$0.13 MHz ($\alpha = 4.21\pm0.02$ dB/m) for Sample 2.
%error bars indicate standard error of the mean.
%The 3.25 MHz difference is statistically significant (Welch t-test, p < 0.001) and unexpected, given identical fabrication conditions.
%The 3.25 MHz difference is unexpected, given identical fabrication conditions.
Kerr-calibrated thermal response measurement reveals a 1.9-fold stronger thermal absorption loss in Sample 2 (Fig. \ref{Fig:2}b, see Method), accounting for their overall loss difference (Fig. \ref{Fig:2}a).
Sample 1 exhibits a long soliton 'step' (330 MHz), which enables DKS generation with simple piezo tuning, while Sample 2 shows a 38-fold reduction in soliton 'step' length (8.6 MHz), precluding access to the DKS state using the same method (Fig. \ref{Fig:2}c).
Similar DKS inaccessibility to Sample 2 has been observed in prior studies, even in high-quality-factor microresonators ($Q \sim 25.6 \times 10^6$)~\cite{liu2024fabrication}.
%SIMS measurement unveils a 10-fold higher Cu concentration in Sample 2, consistent with the stronger thermal effect.

%Figure \ref{Fig:2}d shows depth-resolved Cu concentration profiles in the two samples.
SIMS measurement in Figure \ref{Fig:2}d shows depth-resolved Cu concentration profiles in the two samples.
%	Notably, Cu was below the detection limit in the SiO$_2$ cladding layers (depth$<$ 0.3 $\mu$m and $>$ 1.5 $\mu$m) but significant in the Si$_3$N$_4$ waveguide layer.
	Corrected Cu concentrations, accounting for the Si$_3$N$_4$/SiO$_2$ filling ratio, were $\sim1\times10^{16}$ atoms/cm$^3$ in Sample 1 ($\sim$138 ppba, Methods) and $\sim1\times10^{17}$ atoms/cm$^3$ in Sample 2 ($\sim$1380 ppba).
	This tenfold higher Cu concentration in Sample 2 accounts for its increased propagation loss, thermal absorption, and soliton suppression.
%	From Cu concentration and loss variation, the estimated Cu-induced loss in Si$_3$N$_4$ is $\sim$0.52$\pm$0.03 dB/m/ppm, higher than that of Cu$^{2+}$ in fluoride fiber (0.21 dB/m/ppm, Figure \ref{Fig:1}f), suggesting a stronger interaction between Cu and the Si$_3$N$_4$ host material (Supplementary Note 3).
We note that the estimated Cu-induced loss in Si$_3$N$_4$ is $\sim0.52\pm0.03$ dB/m/ppm, higher than that of Cu$^{2+}$ in fluoride fiber (0.21 dB/m/ppm, Figure \ref{Fig:1}f), suggesting a stronger interaction between Cu and the Si$_3$N$_4$ host material (Supplementary Note 3).

In a separate experiment, we intentionally contaminated a Si$_3$N$_4$ microresonator with Cu by diffusion (Supplementary Note 4).
The original 620-MHz-long soliton 'step' was reduced by a factor of 62, accompanied by an extended MI region and an increased thermal buildup.
The thermal absorption loss also increased by a factor of 4.4.
This experiment serves as an additional proof that indeed Cu impurities cause the thermal effects that hinder DKS accessibility.

\subsection*{Copper Impurities: Origins from Si Wafers}

To understand the origin of Cu impurities in Si$_3$N$_4$ waveguides, we conducted SIMS analysis after each fabrication step in the photonic Damascene process \cite{liu_high-yield_2021} (Supplementary Note 5).
The possibility of Cu contamination from most fabrication steps and their associated equipment has been ruled out, e.g., reactive ion etching (RIE) tools and low-pressure chemical vapor deposition (LPCVD) furnaces. However, significant Cu contamination appeared after the 1200$^{\circ}$C annealing step, which is necessary to reduce losses from Si-H and N-H bonds.
%Since no evidence of Cu was found in the annealing furnaces (and Cu concentration variations among devices processed in the same tube contradicted this hypothesis), the only remaining suspect of Cu contamination source is the silicon wafer. This is, however, in contrary to the null detection level of Cu in commercial electronic-grade silicon wafer surfaces.
As no evidence of Cu was found in the annealing furnaces (and device-to-device variations in Fig.\ref{Fig:2} contradict this hypothesis), the Si wafers remain the only suspected source of contamination.
This is, however, contrary to the low surface contamination specified by the manufacturer and the non-detectable Cu levels in commercial EG-Si wafers.
%Si$_3$N$_4$ is a well-known material that can getter transition metals from Si, including Fe and Cu, at high temperatures~\cite{myers2000mechanisms,liu2016gettering,inglese2018cu,liu_gettering_2019,le2021impurity}. 
%In this process, Cu can diffuse from Si into and aggregating in Si$_3$N$_4$ during annealing.
%Although commercial specifications ensure low surface Cu levels in Si,
%Cu can evade detection due to its high diffusivity~\cite{istratov1998intrinsic}, which results in uniform dilution in the Si bulk below SIMS sensitivity limits.

Si$_3$N$_4$ is known to getter transition metals such as Fe and Cu from Si wafers~\cite{myers2000mechanisms,liu2016gettering,inglese2018cu,liu_gettering_2019,le2021impurity}, with impurities diffusing into and aggregating within Si$_3$N$_4$ at high temperatures.
%Therefore, during the 1200 $^{\circ}$C annealing step in Si$_3$N$_4$ PIC fabrication, diluted Cu from the Si substrate diffuses and accumulates in the Si$_3$N$_4$ layer, enabling detection by SIMS.
%In addition, the lack of high-temperature processing (e.g., $>$500$^{\circ}$C) in standard CMOS fabrication prevents Cu from reaching the device, likely explaining why this issue remained undetected until now.
%We note that the Cu concentrations in Samples 1 and 2 are consistent with those in their respective Si wafers, when using a dedicated Si$_3$N$_4$ gettering layer (Fig.~\ref{Fig:2}e).
%This identifies the Si substrate as the source of Cu impurities in Si$_3$N$_4$ waveguides.
%By annealing Si$_3$N$_4$ thin films at high temperature, Cu concentration in bulk silicon wafers can reduce by orders of magnitude.
On the other hand, Cu exhibits high diffusivities in Si and SiO$_2$ \cite{kim2006extraction}: it can diffuse through a 525-$\mu m$-thick Si wafer in approximately 3 hours at room temperature, or within minutes at elevated temperatures ($>$ 900 $^{\circ}$C) (Supplementary Note 2). 
In EG-Si wafers, Cu can evade detection due to its high diffusivity, resulting in uniform dilution in bulk Si wafers below SIMS sensitivity limits.
However, during the 1200$^{\circ}$C annealing process in Si$_3$N$_4$ PIC fabrication, the dilute Cu impurities from bulk Si diffuse into and are gettered in the Si$_3$N$_4$ layer, explaining the detectable Cu concentration by SIMS.
We note that the Cu concentrations observed in Sample 1 and Sample 2 are in fact consistent with those in their respective Si wafers, when using a dedicated Si$_3$N$_4$ gettering layer (Fig. \ref{Fig:2}e).

To further prove that Cu impurities indeed come from Si, we annealed three Si wafers (with identical initial Cu levels) coated with 200 nm plasma-enhanced chemical vapor deposition (PECVD) Si$_3$N$_4$ at 800 $^{\circ}$C (Fig. \ref{Fig:2}f). 
The depth profiles of Cu concentration (Fig. \ref{Fig:2}g) show progressive accumulation in Si$_3$N$_4$ over annealing time of 0, 4, and 10 hours.
Crucially, Cu impurities were absent in the as-deposited PECVD Si$_3$N$_4$, excluding the Si$_3$N$_4$ film as the contamination source.
These findings confirm that Cu impurities originate from bulk Si wafers, initially accumulating at the Si/Si$_3$N$_4$ interface and  diffusing into Si$_3$N$_4$ until saturation is reached during annealing.

\begin{figure*}[t!]
		\centering
		\includegraphics[width=\textwidth]{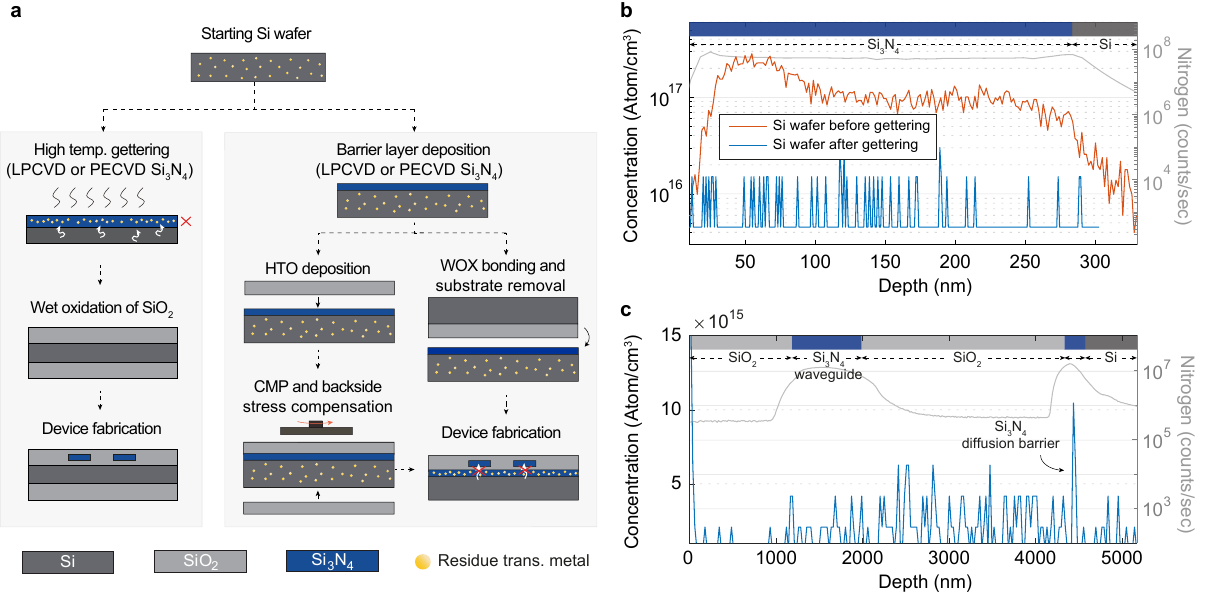}
		\caption{ 
			\footnotesize
			\textbf{Substrate preparation techniques for copper-impurity-free PICs.} 
%			\textbf{Methods for Cu-free PIC fabrication for deterministic DKS generation.} 
			\textbf{a,} Schematic representation of two fabrication approaches.
			The left panel illustrates direct gettering for Si wafer purification, where a defective LPCVD or PECVD Si$_3$N$_4$ layer ($\sim$ 200 nm) is deposited on Si wafers without native SiO$_2$.
			Cu impurities from the Si wafer diffuse into the Si$_3$N$_4$ layer during high-temperature annealing (800$^{\circ}$C, 10 hours), and the contaminated layer is subsequently removed by dry and wet etching.
			The right panel depicts a Cu diffusion barrier method, where a 200-nm LPCVD Si$_3$N$_4$ layer is deposited directly on Si wafers to prevent Cu migration into the device layer.
			\textbf{b,} Cu depth profile in Si wafers before and after the direct gettering process, showing near-complete Cu removal.
			\textbf{c,} Cu depth profile in Si$_3$N$_4$ devices with a diffusion barrier, revealing a Cu peak ($\sim$ 10$^{16}$ atoms/cm$^3$) in the barrier layer, while Cu remains at the detection limit in the waveguide layer.
			}
		\label{Fig:3}
	\end{figure*}

The wafer-to-wafer Cu concentration variations directly impact photonic device performance, explaining the observed differences in propagation loss, thermal absorption, and soliton generation efficiency, elucidating the consistent ease of soliton generation in our prior studies \cite{liu2018ultralow, riemensberger_massively_2020, shen2020integrated, liu2020photonic, liu_high-yield_2021}.
	
\subsection*{Copper Gettering: Substrate Preparation for Copper-Impurity-Free PICs}
	
%After identifying Cu contamination originating from the bulk of CMOS-grade Si wafers as a key loss mechanism in Si$_3$N$_4$ waveguides, we propose two methods to prepare impurity-free Si substrates for ultra-low loss photonic integrated circuits.
After attributing the dominant thermal effect in Si$_3$N$_4$ waveguides to Cu impurities originating from CMOS-grade Si wafers, we develop two methods to produce copper-impurity-free substrates for ultra-low-loss Si$_3$N$_4$ PICs.
%After tracing waveguide thermal losses to bulk Cu in CMOS-grade Si wafers, we develop two methods to produce impurity-free substrates for ultra-low-loss Si$_3$N$_4$ PICs.
%Both approaches, illustrated in Fig.~\ref{Fig:3}a, utilize Si$_3$N$_4$ layers to suppress impurity diffusion during high-temperature processing.

The first method (Fig.~\ref{Fig:3}a, left) employs a sacrificial Si$_3$N$_4$ gettering layer deposited via LPCVD or PECVD (see Supplementary Note 6 for comparison).
Native SiO$_2$ is removed by HF etching prior to Si$_3$N$_4$ deposition.
During a subsequent high-temperature anneal (800 $^{\circ}$C, 10 h), Cu impurities migrate to the Si$_3$N$_4$ layer, where they are trapped by its high solubility, reduced impurity mobility, and defect-mediated gettering \cite{myers2000mechanisms,liu_gettering_2019}.
%The cooling rate critically determines gettering efficiency, as demonstrated by temperature-dependent Cu segregation kinetics in \cite{inglese2018cu}:
%Controlled slow cooling ($-$4$^{\circ}$C/min, 800-600$^{\circ}$C) maximizes gettering effectiveness by maintaining an optimal solubility gradient between the gettering layer and the bulk, while maintaining sufficient Cu diffusivity for near-complete impurity segregation.
%Conversely, rapid air-cooling ($\sim -$240$^{\circ}$C/min) produces insufficient segregation, leaving residual bulk Cu due to kinetically limited diffusion.
The cooling rate governs gettering efficiency \cite{inglese2018cu}: slow cooling (e.g., $-$4$^{\circ}$C/min) optimizes the solubility gradient and Cu diffusivity, enabling near-complete impurity segregation (Supplementary Note 7).
Following annealing, the Cu-enriched Si$_3$N$_4$ is stripped, physically removing contaminants (surface characterization and oxidation detailed in Supplementary Note 8).
%A smooth SiO$_2$ cladding layer is then formed via wet oxidation and chemical-mechanical polishing, enabling fabrication of photonic structures on purified substrates.
SIMS analysis in Fig.~\ref{Fig:3}b confirms near-complete Cu removal in gettered Si wafers,
%with Cu concentrations reduced to the detection limit, 
demonstrating the efficacy of the gettering process.
%The characterization and oxidation of purified Si wafers are detailed in Supplementary Note 10.

The second method (Fig.~\ref{Fig:3}a, right) integrates a 200-nm Si$_3$N$_4$ diffusion barrier within the substrate.
This LPCVD-deposited buried layer simultaneously getters Cu impurities during Si$_3$N$_4$ waveguide annealing and blocks their upward diffusion into the waveguide core.
LPCVD is selected over PECVD to avoid issues such as degassing or defect formation during high-temperature processes.
SIMS depth profiling in Fig.~\ref{Fig:3}c reveals Cu concentrations below the detection limit in the Si$_3$N$_4$ waveguide layer, while the barrier exhibits a pronounced accumulation peak ($>10^{16}$~cm$^{-3}$).
This approach ensures minimal Cu diffusion into optically active regions while maintaining compatibility with multilayer integration.

Supplementary Note 9 demonstrates that Si$_3$N$_4$ microresonators fabricated on Cu-gettered Si wafers exhibit superior performance metrics compared to non-gettered counterparts: reduced propagation loss (by 1.53 dB/m), weaker thermal absorption, complete suppression of Cu contamination, and extended soliton step lengths.
These results validate the efficacy of our gettering approach for enabling ultra-low-loss PICs.
%which facilitate high-performance applications, including deterministic soliton generation.

	\subsection*{Deterministic Soliton Microcomb Generation}	
	
	\begin{figure*}[t!]
		\centering
		\includegraphics[width=\textwidth]{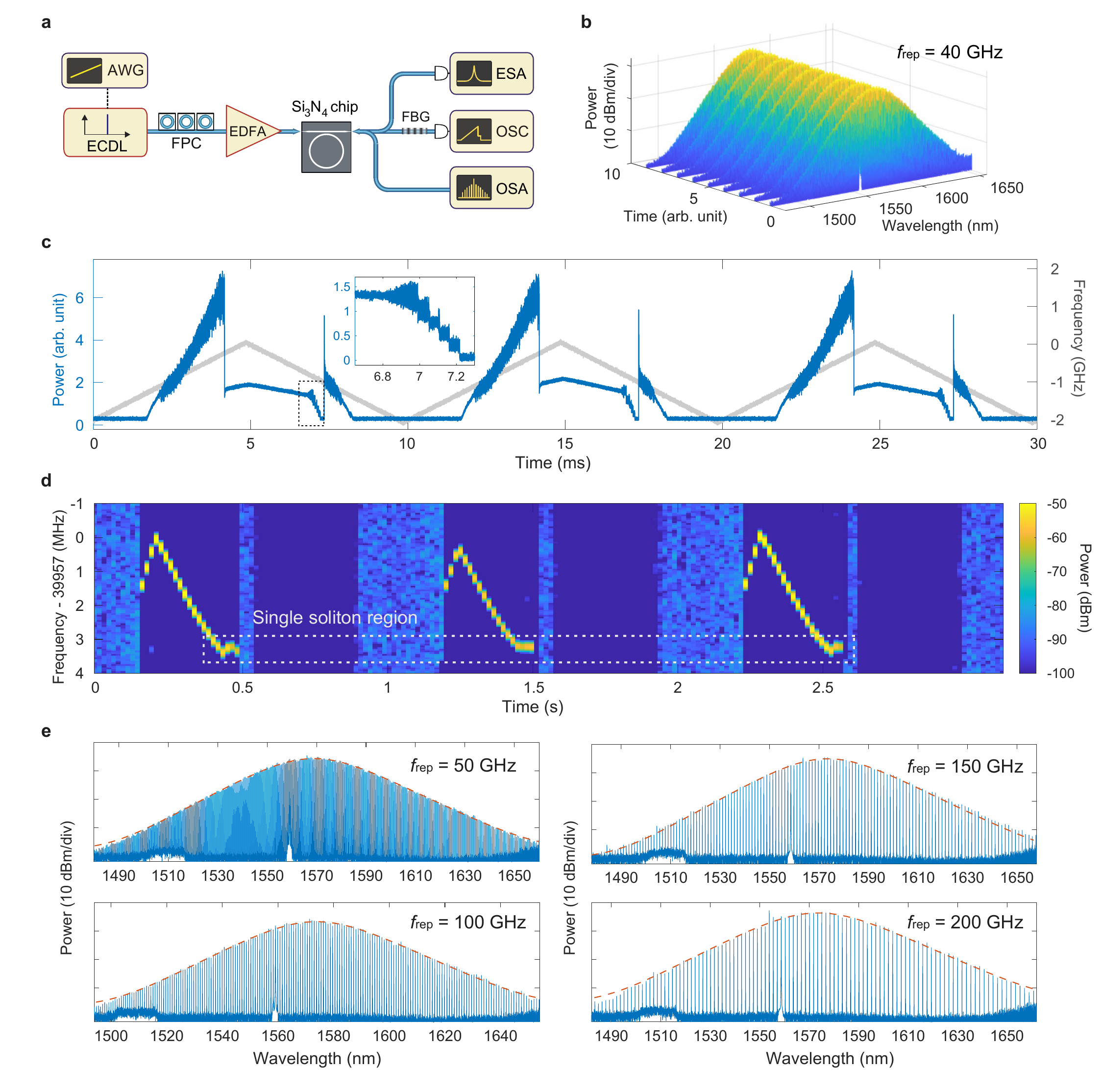}
		\caption{
			\footnotesize
			\textbf{Deterministic DKS generation in Si$_3$N$_4$ microresonators fabricated with copper-impurity-free Si substrates using the photonic Damascene process.}
			\textbf{a,} Experimental setup for DKS generation. ECDL: external-cavity diode laser; AWG: arbitrary waveform generator; FPC: fiber polarization controller; EDFA: erbium-doped fiber amplifier; FBG: fiber Bragg grating; ESA: electrical spectrum analyzer; OSC: oscilloscope; OSA: optical spectrum analyzer.
			\textbf{b,} Single soliton spectra from 9 consecutive measurements in a 40-GHz FSR microresonator, with pump wavelengths filtered for clarity.
%			Device ID: \texttt{D163\_01\_F4\_C12}.
Device ID: \texttt{D163\_03\_F4\_C12}.
			\textbf{c,} Soliton steps observed during a continuous wavelength sweep across the microresonator resonance.
			The inset presents a zoomed-in view of the backward tuning regime, showing successive transitions from 5 solitons to 0 solitons.
			\textbf{d,} Soliton repetition rate recorded via photodetection and an electrical spectrum analyzer during continuous wavelength sweeping.
			\textbf{e,} Single soliton spectra from four microresonators with FSRs of 50 GHz, 100 GHz, 150 GHz, and 200 GHz, respectively, fabricated on Cu-gettered Si wafers.
			All microresonators have a cross-section of 2.2$\times$0.9~$\mu$m$^2$ and are pumped with TE-polarized light.
			Pump wavelengths are filtered for clarity.
			Device IDs: \texttt{D163\_03\_F1\_C2} (50~GHz), \texttt{D163\_03\_F1\_C3} (100~GHz), \texttt{D163\_03\_F1\_C6} (150~GHz), and \texttt{D163\_03\_F1\_C7} (200~GHz).}
		\label{Fig:4}
	\end{figure*}
	
We next demonstrate soliton generation using Si$_3$N$_4$ microresonators fabricated on copper-impurity-free Si substrates. Without employing complex DKS stabilization techniques such as auxiliary lasers, active feedback, or pulse pumping, we achieved a 100$\%$ success rate in generating DKSs across all Si$_3$N$_4$ microresonator chips. Single-soliton states were reliably accessed with simple piezo tuning, as evidenced by nine consecutive measurements in a 40-GHz-FSR microresonator (Fig. \ref{Fig:4}b). DKS formation dynamics were monitored on an oscilloscope (Fig. \ref{Fig:4}c). By slowly ramping the pump wavelength (at frequencies $<$100 Hz, well below the thermal response frequency of kHz to MHz), distinct stages of DKS formation, including primary combs, MI states, and soliton states, were repeatably observed. Notably, backward tuning enabled deterministic soliton state switching, reducing the number of solitons sequentially \cite{guo_universal_2017}.
These dynamics were further corroborated by the repetition rate measurements, recorded via a photodiode and electrical spectral analyzer, with high signal-to-noise-ratio signals exclusively appearing during DKS formation (Fig. \ref{Fig:4}d). 

We present a gallery of soliton microcombs with repetition rates spanning 50 GHz to 200 GHz (Fig. \ref{Fig:4}e). The FSR lower limit is constrained by our design; however, 10 GHz DKSs previously demonstrated with the standard photonic Damascene process \cite{liu2020photonic} are attainable with higher reliability using the copper-impurity-free process.
Furthermore, the extended soliton steps of these microresonators enable direct access to soliton microcombs through manual tuning of the pump laser's piezoelectric actuator, eliminating the need for arbitrary waveform generators.
This simplification reduces the complexity of DKS generation setups, paving the way for broader adoption and application of soliton microcombs.
	
\subsection*{Summary}
We identified Cu impurities as the primary cause of thermal absorption and soliton inaccessibility in ultra-low-loss Si$_3$N$_4$ photonic circuits.
Through systematic SIMS analysis and thermal response measurements, we demonstrate the surprising finding that Cu impurities originate from the bulk of CMOS-grade Si wafers and diffuse into Si$_3$N$_4$ waveguides during high-temperature annealing, degrading device performance.
By developing two robust Cu-gettering techniques, one with direct gettering and the other with a diffusion barrier, we reduce thermal effects and achieve copper-impurity-free Si$_3$N$_4$ PICs, enabling deterministic DKS generation with repetition rates as low as 40 GHz, a regime that has traditionally been challenging.
%This advancement facilitates high-performance applications such as turnkey soliton generation, supercontinuum generation, and traveling-wave parametric amplification.
Furthermore, our substrate preparation methods are fully compatible with commercial foundry processes, allowing immediate integration into standard production workflows.
This work not only addresses a longstanding challenge hindering the widespread adoption of microcomb technology, but also paves the way for the broad deployment of ultra-low-loss photonic integrated circuits in optical communications, sensing, computing, and quantum technologies.
%We identify copper as the dominant source of residual thermal absorption in ultra-low-loss Si$_3$N$_4$ photonic integrated circuits, revealing that Cu impurities—originating from the bulk of CMOS-grade silicon wafers—migrate into the waveguide core during high-temperature processing. This discovery mirrors the pivotal finding in the 1960s that transition metal impurities, particularly Cu, were responsible for optical losses in early fiber optics—a breakthrough that enabled the development of modern optical communication.
%Our work represents a similarly foundational advance for integrated photonics: by uncovering and addressing Cu-induced absorption, we remove a long-standing barrier to deterministic dissipative Kerr soliton (DKS) generation. Through wafer-level Cu removal techniques, we demonstrate turnkey soliton microcombs with wide soliton steps and repeatable single-soliton access—without complex tuning protocols.
%Beyond enabling robust DKS generation, our Cu mitigation methods open a path toward even lower propagation loss in Si$_3$N$_4$ PICs. Crucially, the process is fully compatible with commercial foundry workflows, paving the way for the widespread deployment of microcombs in applications spanning optical synthesis, sensing, computing, and quantum technologies.
	\pretolerance=0
	\bigskip
	
%\section*{Methods}
\noindent
\textbf{Methods}
\medskip

\begin{footnotesize}
\noindent \textbf{Conversion of impurity concentration from atoms/cm\(^3\) to ppba}:
The total atomic density $N_{\text{total}}$ (in cm$^{-3}$) of a pure Si$_3$N$_4$ film, is calculated using the molecular density $\rho_{\text{Si}_3\text{N}_4}$, molar mass $M_{\text{Si}_3\text{N}_4}$, and Avogadro's number:

\[
N_{\text{total}} = \frac{\rho_{\text{Si}_3\text{N}_4} \cdot N_A}{M_{\text{Si}_3\text{N}_4}}\times 7 \approx 9.53 \times 10^{22} \, \text{atoms/cm}^3
\]
where:
\(\rho_{\text{Si}_3\text{N}_4} = 3.17 \, \text{g/cm}^3\) (density of Si\(_3\)N\(_4\)), and \(N_A = 6.022 \times 10^{23} \, \text{atoms/mol}\) (Avogadro's number), \(M_{\text{Si}_3\text{N}_4} = 140.28 \, \text{g/mol}\) (molar mass of Si\(_3\)N\(_4\)).

The ppba value is calculated by expressing the impurity concentration, \(C_{\text{Cu}}\) (in cm$^{-3}$), as a fraction of \(N_{\text{total}}\) and multiplying by \(10^9\):
\[
\text{ppba} = \frac{C_{\text{Cu}}}{N_{\text{total}}} \times 10^9
\]

For instance, \(C_{\text{Cu}} = 10^{16} \, \text{atoms/cm}^3\) corresponds to a ppba value of:
\[
\text{ppba} = \frac{10^{16}}{9.53 \times 10^{22}} \times 10^9 \approx 105
\]

In a mixed layer composed of e.g. 21.76$\%$ Si$_3$N$_4$ and 78.24$\%$ SiO$_2$, the total atomic density of the material must be calculated as a weighted average of the atomic densities of Si$_3$N$_4$ and SiO$_2$.
For SiO$_2$, $\rho_{\text{Si}\text{O}_2}$ = 2.2 g/cm$^3$, $M_{\text{Si}\text{O}_2}$ = 60.08 g/mol:
\[
N_{\text{SiO}_2} = \frac{2.2 \times 6.022 \times 10^{23}}{60.08}\times 3 \approx 6.62 \times 10^{22} \, \text{atoms/cm}^3
\]

The total atomic density of the mixed layer is calculated as:
\[
N_{\text{mixed}} = P_{\text{Si}_3\text{N}_4} \cdot N_{\text{Si}_3\text{N}_4} + P_{\text{SiO}_2} \cdot N_{\text{SiO}_2} \approx  7.25\times 10^{22} \, \text{atoms/cm}^3
\]
where $P_{\text{Si}_3\text{N}_4} = 0.2176$ and $P_{\text{SiO}_2} = 0.7824$.

%For a Cu$^+$ concentration of 1$\times$10$^{16}$ atoms/cm$^3$ in the %Si$_3$N$_4$/SiO$_2$ mixed layer, the concentration in ppba is converted as:
%	\[
%	\text{ppba} = \frac{C_{\text{Cu}}}{N_{\text{mixed}}} \times 10^9 \approx 501.88
%	\]

\bigskip
\noindent \textbf{Thermal response measurement of Si$_3$N$_4$ microresonators}:
	To quantify thermal absorption loss $\kappa_\mathrm{abs}/2\pi$ in Si$_3$N$_4$ microresonators, we measure the resonance frequency shift of a probe mode induced by intensity-modulating a pump mode.
The pump laser, using low power to maintain linear operation, is tuned to the target resonance and modulated from 1 kHz to 200 MHz, while the probe laser is loosely locked to a probe resonance.
The modulation alters the intracavity photon number, inducing a probe mode frequency shift through thermal and Kerr effects:
	\[\chi(\omega) = \chi_\mathrm{therm}(\omega) + \chi_\mathrm{Kerr}(\omega)\]
where $\chi_\mathrm{therm}(\omega)$ and $\chi_\mathrm{Kerr}(\omega)$ can be distinctly resolved, with $\chi_\mathrm{therm}(\omega)$ dominating at lower frequencies and $\chi_\mathrm{Kerr}(\omega)$ at higher frequencies.
Using the fitted values of $\chi_\mathrm{therm}(\omega)$ and $\chi_\mathrm{Kerr}(\omega)$ at DC ($\omega = 0$), the absorption rate is calculated as:
	\[\kappa_\mathrm{abs} = \frac{2c\nu n_2}{n_\mathrm{g}n_\mathrm{eff}V(d\nu /dP_\mathrm{abs})}\gamma\]
	where $V$ is the optical mode volume, $n_\mathrm{g}$ is the group index, $n_\mathrm{eff}$ is the effective index, and $d\nu /dP_\mathrm{abs}$ is the photothermal frequency shift gradient at pump frequency $\nu$, mainly determined by the material thermo-optic coefficients and is calculated from simulation. The response ratio $\gamma$ is the DC offset of the measurement response function $\gamma(\omega)$ in the Fourier domain, and is obtained through fitting the measured response function. By evaluating the response ratio $\gamma = \chi_\mathrm{therm}(\omega=0)/\chi_\mathrm{Kerr}(\omega=0)$, we can infer the thermal losses from different Si$_3$N$_4$ microresonators.

\bigskip
\noindent \textbf{Soliton generation in impurity-free Si$_3$N$_4$ microresonators}:
An ECDL is amplified by an EDFA and coupled to a Si$_3$N$_4$ microresonator chip. The ECDL wavelength is tuned via a piezoelectric unit controlled either manually or by an AWG.
To generate soliton microcombs, the pump laser is scanned with decreasing optical frequency across a resonance of the microresonator, passing through MI states and finally into the soliton states in the effective red-detuned regime. For microresonators with significant thermal effects, the MI-to-soliton transition induces a temperature drop and resonance shift, potentially destabilizing soliton states if the soliton step is smaller than the effective detuning. 
In contrast, impurity-free Si$_3$N$_4$ microresonators with small thermal effects exhibit deterministic soliton generation due to long soliton steps and marginal intracavity temperature changes. 
To reach a single-soliton state after stably landing on multi-soliton states, the pump laser frequency is tuned backwards with increasing frequency, leading to the successive extinction of intracavity solitons down to the single-soliton state \cite{guo_universal_2017}. 

\bigskip
\noindent \textbf{Funding Information}: This work was supported by the Horizon Europe EIC transition programme under grant No. 101136978 (CombTools), and funded by the Swiss State Secretariat for Education, Research and lnnovation (SERI). This work was further supported by the Swiss National Science Foundation under grant agreement No. 216493 (HEROIC).

\noindent \textbf{Acknowledgments}:
The Si$_3$N$_4$ samples were fabricated in the EPFL Center of MicroNanoTechnology (CMi).
The SIMS measurements were performed by Dr. Xi Liu at Eurofins EAG Laboratories.

\noindent \textbf{Author contribution}:
T.J.K. conceived the idea and concept.
X.J. designed the Si$_3$N$_4$ samples.
X.J., X.L., R.N.W., and Z.Q. fabricated the Si$_3$N$_4$ samples.
X.J. designed and performed the Cu gettering experiment with assistance from Z.Q..
X.J. and X.L. characterized and analyzed the Si$_3$N$_4$ samples.
X.L. performed the soliton generation experiment with assistance from G.L..
X.J. and X.L. wrote the manuscript with input from all co-authors.
T.J.K. supervised the project.

\noindent \textbf{Data Availability Statement}: The code and data used to produce the plots within this work will be released on the repository \texttt{Zenodo} upon publication of this preprint.

\noindent \textbf{Competing interests}: T.J.K. is a cofounder and shareholder of LiGenTec SA, a start-up company offering Si$_3$N$_4$ photonic integrated circuits as a foundry service.

\end{footnotesize}
\bibliographystyle{apsrev4-2}
\bibliography{bibliography_manual_v2}

\end{document}

	%\includepdf[landscape=false]{Science_SM_Cover.pdf}
	\title{Supplementary Information for: Copper-impurity-free photonic integrated circuits enable deterministic soliton microcombs}
	
	\author{Xinru Ji$^{1,2}$, 
		Xurong Li$^{1,2}$,
		Zheru Qiu$^{1,2}$,
		Rui Ning Wang$^{3}$,
		Marta Divall$^{1,2}$,
		Andrey Gelash$^{1,2}$,
		Grigory Lihachev$^{1,2}$,
		and Tobias J. Kippenberg$^{1,2\dag}$}
	\affiliation{
		$^1$Institute of Physics, Swiss Federal Institute of Technology Lausanne (EPFL), CH-1015 Lausanne, Switzerland\\
		$^2$Institute of Electrical and Micro Engineering, Swiss Federal Institute of Technology Lausanne (EPFL), CH-1015 Lausanne, Switzerland\\
		$^3$Luxtelligence SA, CH-1015 Lausanne, Switzerland}
	
\setcounter{equation}{0}
\setcounter{figure}{0}
\setcounter{table}{0}

\setcounter{subsection}{0}
\setcounter{section}{0}
\setcounter{secnumdepth}{3}
	
	\begin{abstract}
		Supplementary Information for this manuscript includes: Si$_3$N$_4$ numerical simulations of thermal effects on soliton generation, detection of metal impurities in Si$_3$N$_4$ waveguides, analysis of Cu diffusion in Si and absorption features in Si$_3$N$_4$, deliberate Cu contamination experiments and their impact on soliton formation, pathways of Cu diffusion into Si$_3$N$_4$ waveguides, gettering efficiency comparison between LPCVD and PECVD Si$_3$N$_4$, discussion of diverse gettering approaches, characterization of purified Si wafers post-gettering, characterization of Si$_3$N$_4$ devices fabricated from gettered Si wafers, analysis of Cu impurities in commercial Si$_3$N$_4$ photonic integrated circuits and in LNOI/LTOI platforms.
	\end{abstract}
	
	\maketitle
{\hypersetup{linkcolor=blue}\tableofcontents}
\newpage

%%%
	%%%%%%%%%%%%%%%%%%%%%%%%%%%%%%%%%%%%%%%%%%%%%%%%%%%%%%%%%%%%%%%
	\section{Numerical simulations of thermal effects on soliton generation}
	To illustrate the role of thermal effects on soliton generation as shown in the main text, we simulate the normalized Lugiato–Lefever equation (LLE) coupled to a cavity temperature equation:
	\[ i \frac{\partial \Psi}{\partial \tau} + \frac{1}{2} \frac{\partial^2 \Psi}{\partial \theta^2} + |\Psi|^2\Psi = (-i + \zeta_{\mathrm{abs}} + \Theta)\Psi +if,  \]
	\[ \frac{\partial \Theta}{\partial \tau} = \frac{2}{\kappa \tau_T} \biggl(\frac{n_{2T}}{2\pi n_2} \int_0^{2\pi} |\Psi|^2 d\theta - \Theta \biggr). \]
	Here, we use the following dimensionless variables: the slow time $\tau$, the internal cavity coordinate $\theta$, the absolute cavity detuning $\zeta_{\mathrm{abs}}$, variation of the cavity temperature $\Theta$ and the internal cavity field $\Psi(\tau,\theta)$.
	
	In addition, the thermal response is controlled by the thermal relaxation time $\tau_T$ and the thermal nonlinearity coefficient $n_{2T}$.
	The simulation was performed with the standard pseudo-spectral split-step method and parameters taken from \cite{guo_universal_2017}.

\section{Metal impurities in Si$_3$N$_4$ PICs}

\begin{figure*}[h!]
		\centering
		\includegraphics[width=\textwidth]{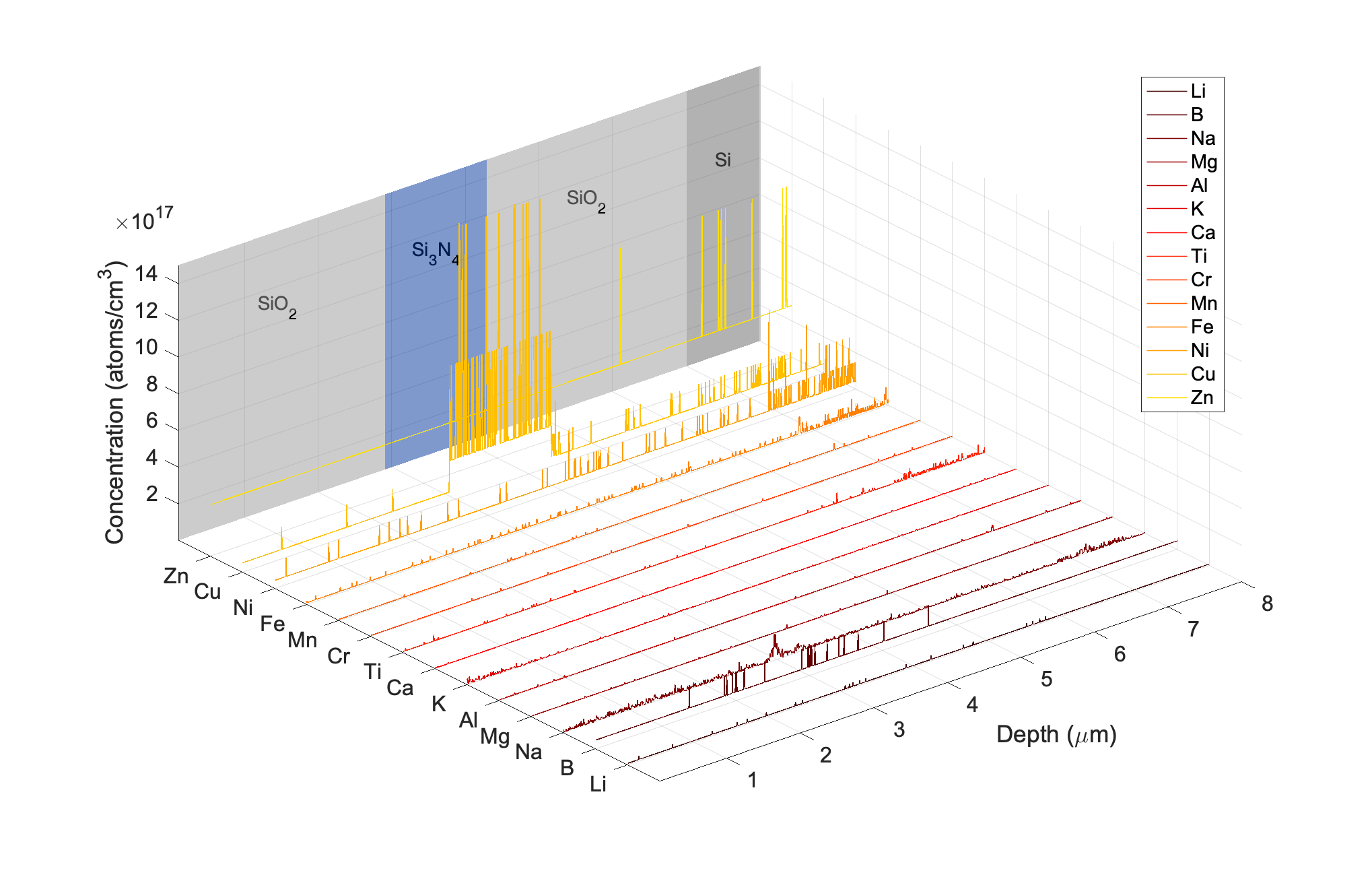}
		\caption{
			\footnotesize
			\textbf{SIMS depth profiles of 14 metal impurities in a fully SiO$_2$-cladded Si$_3$N$_4$ waveguide.}
			The x-axis represents sputtering depth, the y-axis denotes metal species, and the z-axis indicates impurity concentration. Background colors indicate the layer structure. 
			Sample ID: \texttt{D100\_01\_F1\_C7}.
		}
		\label{Fig:SI_metal}
	\end{figure*}

We performed secondary ion mass spectroscopy (SIMS) measurements to analyze metal impurities in a fully SiO$_2$-cladded Si$_3$N$_4$ waveguide.
Secondary ion mass spectroscopy quantifies trace impurities by sputtering the sample with an ion beam and analyzing ejected secondary ions via mass spectrometry, offering parts-per-billion atomic (ppba) sensitivity and depth-resolved profiles.
In SIMS, multiply charged ions (e.g., Cu$^{2+}$) convert to singly charged ions (Cu$^{+}$) via electron capture in the sputtering plasma, so the signal reflects total elemental concentration regardless of initial charge state \cite{vickerman2011surface}.

We investigated metal impurities including lithium (Li), boron (B), sodium (Na), magnesium (Mg), aluminum (Al), potassium (K), calcium (Ca), titanium (Ti), chromium (Cr), manganese (Mn), iron (Fe), nickel (Ni), copper (Cu), and zinc (Zn).
The SIMS depth profiles, presented in Supplementary Figure \ref{Fig:SI_metal}, reveal impurity concentrations across the Si$_3$N$_4$ chip, with background colors indicating the SiO$_2$ claddings, Si$_3$N$_4$ waveguide core, and Si substrate.
Notably, only Cu is detected within the Si$_3$N$_4$ chip, specifically concentrated in the Si$_3$N$_4$ waveguide, while all other metals remain at or below the detection limit.
This finding identifies Cu as the dominant metal impurity in Si$_3$N$_4$ waveguides.

\section{Copper diffusion in Si and absorption features in Si$_3$N$_4$}
	
	\begin{figure*}[h!]
		\centering
		\includegraphics[width=0.9\textwidth]{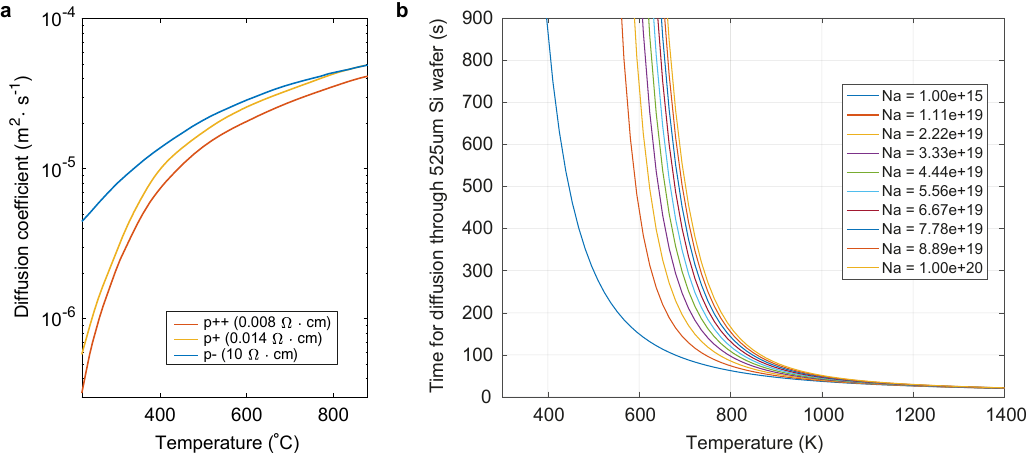}
		\caption{
			\footnotesize
			\textbf{Copper diffusion in doped silicon wafers.} 
			\textbf{a,} Diffusion coefficient of Cu in P/B-doped Si wafers as a function of temperature for varying boron doping concentrations \cite{Hirano_Shabani_2011}.
			\textbf{b,} Time for Cu to diffuse through a 525 $\mu$m Si wafer at increasing boron doping concentrations \cite{istratov2000diffusion}.}
		\label{Fig:SI_Cu_diff}
	\end{figure*}
	
%% Cu diffusion in Si
Copper exhibits high mobility in materials like Si and SiO$_2$ \cite{kim2006extraction}.
They can diffuse through the matrix and be trapped at defect sites or interfaces, forming deep-level defects.
The diffusion behavior of Cu in Si wafers is strongly influenced by doping concentrations, conductivity type, and temperature.
In highly boron-doped p-type Si wafers, Cu exhibits a significantly reduced diffusion coefficient at low temperatures ($<$400$^{\circ}$C), attributed to strong electrostatic interactions between Cu ions and p-type dopants, which impede Cu migration \cite{istratov1997influence,Hirano_Shabani_2011}. 
At elevated temperatures ($>$900$^{\circ}$C), the diffusion coefficients of Cu in p-type wafers converge, as thermal energy overcomes attraction potential, resulting in similar diffusion rates across doping levels (Supplementary Figure \ref{Fig:SI_Cu_diff}a).

The effective diffusion coefficient, as described in \cite{istratov2000diffusion}, is given by:

\begin{equation}
D_{\text{eff}} = \frac{D_{\text{int}}}{1 + \alpha} = \frac{D_{\text{int}}}{1 + \Omega N_{\mathrm{a}}},
\end{equation}

where $N_{\mathrm{a}}$ is the density of trapping centers, and $\Omega$ is the pairing constant, representing the effective capture volume:

\begin{equation}
\Omega = 4 \pi \int_{\Omega_0} r^2 \exp \left[\frac{V(r)}{k_{\mathrm{B}} T}\right] dr.
\end{equation}

$\Omega N_{\mathrm{a}}$ quantifies the ratio of immobilized $\mathrm{Cu}_i^\mathrm{+}$ to mobile $\mathrm{Cu}_i^\mathrm{+}$.

The intrinsic diffusion coefficient, obtained from transient ion drift measurements in \cite{istratov2000diffusion}, is expressed as:

\begin{equation}
D_{\text{eff}}(\mathrm{Cu}) = \frac{3 \times 10^{-4} \times \exp (-2090 / T)}{1 + 2.584 \times 10^{-20} \times \exp (4990 / T) \times (N_{\mathrm{a}} / T)} \quad (\mathrm{cm}^2 \mathrm{~s}^{-1}).
\label{eq:Cu_diff}
\end{equation}

The time $\Delta T$ for Cu to diffuse through a distance $\Delta x$ is calculated as $\Delta T = \Delta x^2 / 2D$. According to Eq. \ref{eq:Cu_diff}, Cu can diffuse through a 0.525 mm intrinsic Si wafer in approximately 3 hours at room temperature, or through a moderately B-doped wafer ($N_{\mathrm{a}}=10^{15}$ cm$^{-3}$, $\sim 13.5~\Omega \cdot$cm) in about 15 hours (Supplementary Figure \ref{Fig:SI_Cu_diff}b).
Consequently, for low-temperature processing applications without substrate resistivity constraints, highly boron doped p-type wafers are preferred due to their superior control of Cu-related defects and contamination.

In optical fibers, Cu$^{2+}$ is more commonly observed because fibers are typically made in oxygen-rich environments (e.g., silica glass, SiO$_2$), which stabilize Cu$^{2+}$.
In Si$_3$N$_4$ PICs, copper predominantly exists in ionic forms, either as Cu$^+$ or Cu$^{2+}$ rather than as neutral Cu atoms, due to its tendency to interact chemically with the surrounding matrix. The specific charge state of copper depends on the local chemical environment: Cu$^+$ is favored in oxygen-deficient or nitrogen-rich conditions, while Cu$^{2+}$ is more stable in electronegative or oxidative environments.
Techniques such as X-ray photoelectron spectroscopy (XPS) and electron energy loss spectroscopy (EELS) can precisely determine the oxidation state of copper, though our Si$_3$N$_4$ samples exhibit sub-detection-limit concentrations.

\begin{figure*}[h!]
		\centering
		\includegraphics[width=0.5\textwidth]{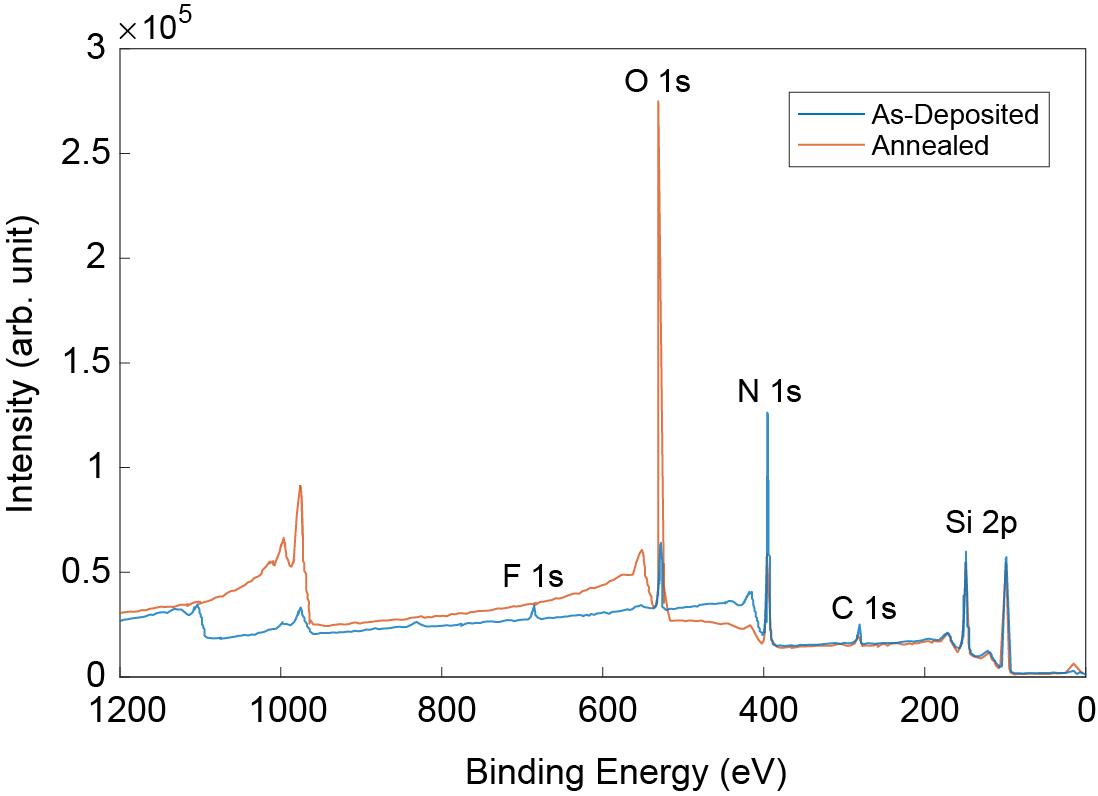}
		\caption{
			\footnotesize
			\textbf{XPS surface analysis of LPCVD Si$_3$N$_4$ films pre- and post-annealing.}
			The post-annealing spectrum shows significant oxygen incorporation, despite the N$_2$ annealing atmosphere.}
		\label{Fig:SI_XPS}
	\end{figure*}

The migration of copper ions in Si$_3$N$_4$ is influenced by the amorphous structure.
Despite the lack of long-range order, silicon and nitrogen atoms form a complex network of closed-loop connections. Within this network, Cu$^\mathrm{+}$ and Cu$^\mathrm{2+}$ ions are postulated to migrate via a "hopping" mechanism between stable interstitial sites formed by these atomic loops under thermal or optical excitation~\cite{zubkov2001modeling}.

The presence of copper ions in Si$_3$N$_4$ can impact thermal absorption losses through ligand-to-metal charge transfer (LMCT) transitions. In LMCT, electrons transition from nitrogen-based ligand orbitals to copper orbitals, generating excited states that non-radiatively relax, converting optical energy into heat \cite{marcus1964chemical,marcus1989relation}. This mechanism dominates over weaker d-d transitions, which are either minimal in Cu$^\mathrm{2+}$ or entirely absent in Cu$^\mathrm{+}$. 

During high-temperature processing (e.g., annealing in N$_2$), the low oxygen content and reducing conditions strongly favor the formation of Cu$^+$. Consequently, Cu in Si$_3$N$_4$ waveguides is expected to exist predominantly as Cu$^+$, with negligible contributions from Cu$^{2+}$ unless oxidizing agents are introduced.
However, XPS surface analysis of Si$_3$N$_4$ films before and after annealing (1200$^{\circ}$C, 11 h, N$_2$) reveals significant oxygen incorporation post-annealing (Supplementary Fig. \ref{Fig:SI_XPS}). This likely stems from residual O$_2$ due to insufficient furnace purging during wafer loading. Although the incorporated oxygen is confined to the surface, it may locally oxidize Cu$^+$ to Cu$^{2+}$ in the waveguides. To mitigate this, annealing in forming gas or under stricter O$_2$-controlled conditions could reduce oxygen infiltration.

\section{Deliberate Cu contamination in Si$_3$N$_4$ microresonators: Impact on DKS generation}

\begin{figure*}[h!]
		\centering
		\includegraphics[width=\textwidth]{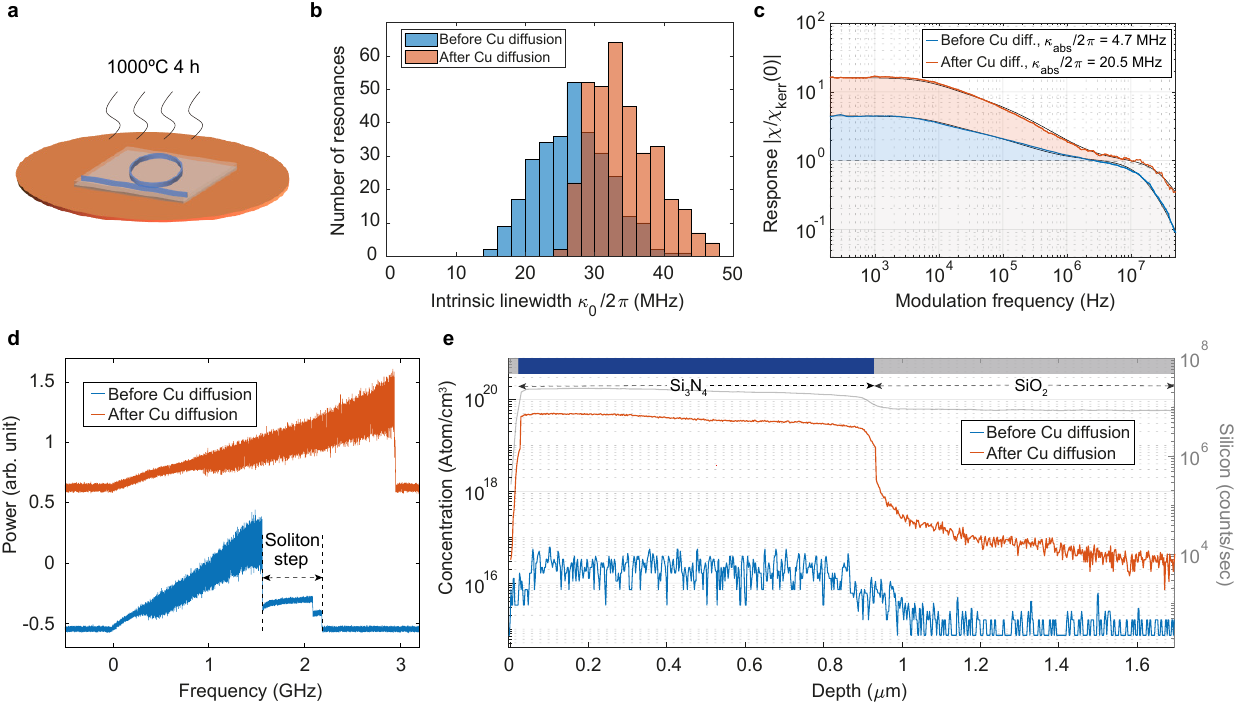}
		\caption{
			\footnotesize
			\textbf{Copper contamination impact on thermal absorption and soliton generation in Si$_3$N$_4$ microresonators.} 
			\textbf{a,} Schematic of the Cu contamination experiment, where a Cu-coated Si wafer is placed beneath the Si$_3$N$_4$ chip during a 1000$^{\circ}$C, 4 hour thermal process in N$_2$.
			Device ID: \texttt{D163\_01\_F2\_C3}.
			\textbf{b,} Histogram of intrinsic loss rate $\kappa_0/2\pi$ before and after Cu diffusion.
			\textbf{c,} Kerr-normalized thermal response measurements before and after deliberate copper diffusion, at the same resonance near 1550~nm in TE polarization.
			\textbf{d,} Measured soliton step lengths in TE polarization, showing shorter steps in microresonators after Cu diffusion.
			\textbf{e,} SIMS depth profiles revealing Cu incorporation into Si$_3$N$_4$ waveguides after thermal treatment.
			SIMS results are calibrated to the Si$_3$N$_4$ filling ratio in the mixed Si$_3$N$_4$/SiO$_2$ waveguide layer.
}
		\label{Fig:SI_Cu_cont}
	\end{figure*}

To further prove that copper introduces additional thermal absorption loss and therefore prevents accessing DKS in microresonators, we deliberately introduced Cu impurities to a Si$_3$N$_4$ microresonator that allowed easy soliton access with low Cu contamination levels.

The Si$_3$N$_4$ chip was placed on a Cu-coated Si wafer, with the bottom Si substrate in direct contact (Supplementary Fig. \ref{Fig:SI_Cu_cont}a).
Cu impurities diffused into the Si$_3$N$_4$ core through the Si substrate and SiO$_2$ bottom cladding during annealing at 1000$^{\circ}$C for 4 hours in a N$_2$ environment.
Statistical analysis of TE$_{00}$ resonances showed a 23$\%$ increase in the most probable intrinsic loss rate $\kappa_0/2\pi$ due to Cu contamination (Supplementary Fig. \ref{Fig:SI_Cu_cont}b).
We characterized the frequency response at the same resonance near 1550 nm for thermal absorption comparison before and after Cu contamination.
The absorption parameter $\gamma$ increased from 3.5 to 15.2, elevating the thermal absorption loss rate $\kappa_\mathrm{abs}/2\pi$ from 4.7 to 20.5 MHz at 1550 nm (Supplementary Fig. \ref{Fig:SI_Cu_cont}c).

In terms of soliton generation, the original long soliton step, measuring 620.33 MHz, shrank to 10.23 MHz after copper diffusion, accompanied by a lengthened MI region, suggesting a profound thermal build-up in Si$_3$N$_4$ microcavity (Supplementary Fig. \ref{Fig:SI_Cu_cont}d).
SIMS measurements (Supplementary Fig. \ref{Fig:SI_Cu_cont}e) show increased Cu impurity concentrations in the Si$_3$N$_4$ layer, confirming contamination via thermal diffusion.
The observed loss increase (in dB/m/ppm) is lower than the Cu$^\mathrm{2+}$-based prediction in \cite{mitachi1983reduction}, as Cu exists in less absorptive states (Cu$^0$ and Cu$^+$) in Si$_3$N$_4$ under inert annealing conditions. 
%Metallic Cu forms nanoparticles that induce optical loss through scattering and absorption, which also degrade DKS generation.

This experiment demonstrates copper impurities as a dominant cause of soliton inaccessibility in ultra-low loss Si$_3$N$_4$ microresonators, highlighting the need to control the Cu impurities when working with ultra-low loss Si$_3$N$_4$ PICs, especially for DKS generation.
%To contrast the effects of Cu absorption, we perform a control experiment where the chip is placed on a bare Si wafer without Cu coating.
%Results show no degradation in soliton steps in the absence of Cu, highlighting the impact of Cu contamination on device performance.

\section{Copper diffusion in Si$_3$N$_4$ waveguides during fabrication processes}

	\begin{figure*}[h!]
		\centering
		\includegraphics[width=\textwidth]{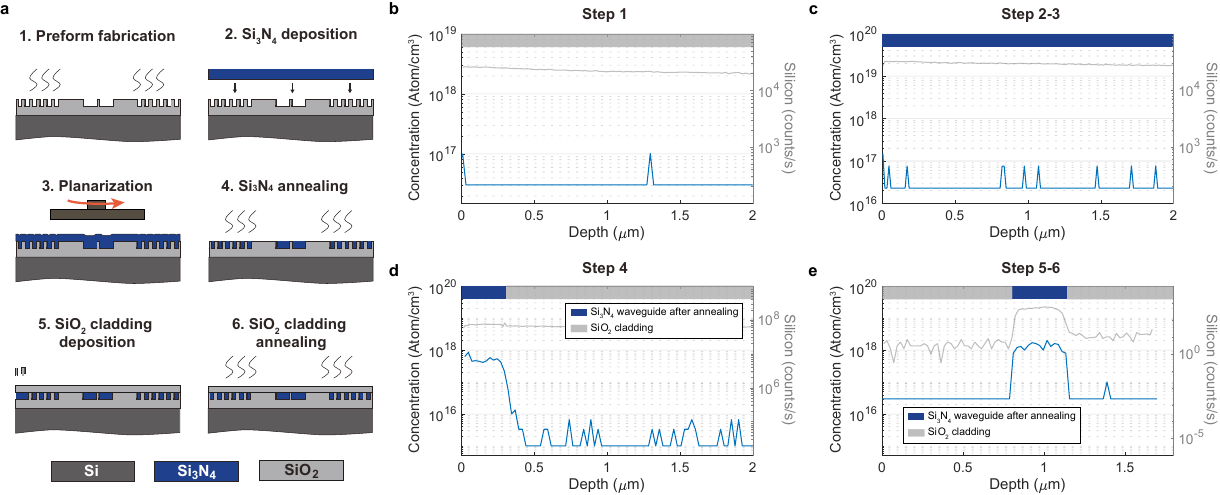}
		\caption{
			\footnotesize
			\textbf{Pathway of Cu diffusion into Si$_3$N$_4$ waveguides during device fabrication.}
			\textbf{a,} Schematic of the photonic Damascene process for low-loss Si$_3$N$_4$ waveguide fabrication, including preform etching, reflow, LPCVD Si$_3$N$_4$ and SiO$_2$ deposition, planarization, and annealing.
			\textbf{b-e,} Depth-resolved Cu concentration profiles: \textbf{b,} after preform etching and reflow; \textbf{c,} post Si$_3$N$_4$ deposition and planarization; \textbf{d,} after Si$_3$N$_4$ annealing (1200 $^{\circ}$C, 11 hours), showing Cu incorporation into the Si$_3$N$_4$ waveguide; \textbf{e,} post SiO$_2$ cladding deposition and annealing, revealing Cu localization exclusively within the Si$_3$N$_4$ waveguide core.}
		\label{Fig:SI_Cu_Dama}
	\end{figure*}

We fabricated Si$_3$N$_4$ waveguides using the photonic Damascene process and employed SIMS to trace the diffusion of Cu impurities throughout the fabrication steps.
The goal was to identify the stage at which Cu incorporates into the Si$_3$N$_4$ waveguide and to determine its origin.
Supplementary Figure \ref{Fig:SI_Cu_Dama}a illustrates the fabrication process including preform etching and reflow (1250$^{\circ}$, 1 hour), low-pressure chemical vapor deposition (LPCVD) of Si$_3$N$_4$, planarization and annealing (1200$^{\circ}$, 11 hour), followed by SiO$_2$ cladding deposition and final annealing (1200$^{\circ}$, 11 hour).

SIMS depth profiles in Supplementary Fig. \ref{Fig:SI_Cu_Dama}b showed no detectable Cu after preform etching and reflow.
Following Si$_3$N$_4$ deposition and planarization, Cu was at barely detectable levels (Supplementary Fig. \ref{Fig:SI_Cu_Dama}c).
After annealing (1200 $^{\circ}$C, 11 hours)-essential for reducing Si-H and N-H bonds in LPCVD Si$_3$N$_4$-Cu concentrations increased significantly within the Si$_3$N$_4$ waveguide core (Supplementary Fig. \ref{Fig:SI_Cu_Dama}d), indicating diffusion during this high-temperature step.
The final profile, after SiO$_2$ cladding deposition and annealing, confirmed Cu localization exclusively within the Si$_3$N$_4$ waveguide, with no detectable Cu in the surrounding SiO$_2$ layers (Supplementary Fig. \ref{Fig:SI_Cu_Dama}e). 

These results demonstrate that Cu diffuses into the Si$_3$N$_4$ waveguide during high-temperature annealing, likely originating from external contamination introduced prior to or during the process.
Experiments in the main text reveal that Si$_3$N$_4$ microresonators fabricated identically on different Si wafers exhibit varying soliton generation behavior, suggesting the Si substrate as the source of Cu impurities.

Although the SIMS measurements were performed on Damascene-processed samples, the observed behavior-Cu incorporation after Si$_3$N$_4$ annealing and its trapping in the Si$_3$N$_4$ core-also applies to subtractive-processed samples, which share different fabrication steps but identical fabrication tools.

\section{Gettering efficiency comparison between LPCVD and PECVD Si$_3$N$_4$}

\begin{figure*}[h!]
		\centering
\includegraphics[width=0.6\textwidth]{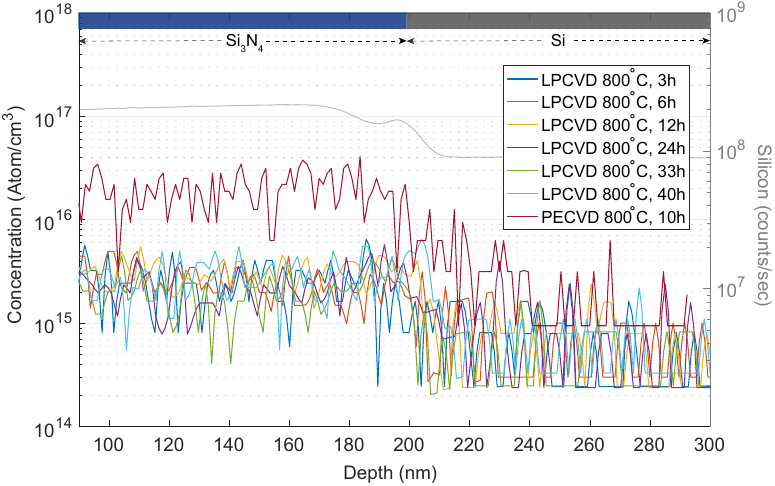}
		\caption{
		\footnotesize
		\textbf{SIMS characterization of Cu concentration in Si wafers coated with LPCVD and PECVD Si$_3$N$_4$ films.}
The deposited Si$_3$N$_4$ films, measuring $\sim$200 nm in thickness for both LPCVD and PECVD, were annealed at 800$^{\circ}$C for varying durations. SIMS analysis reveals higher detectable Cu concentrations in PECVD Si$_3$N$_4$ films compared to LPCVD films, indicating enhanced gettering efficiency in PECVD films due to a higher density of Cu trapping sites. In contrast, LPCVD Si$_3$N$_4$ films exhibit a constant Cu level regardless of annealing time.  
		}
		\label{Fig:SI_LPCVD}
	\end{figure*}

Si$_3$N$_4$ films serve as effective gettering layers, with performance dependent on deposition method. We compared Cu gettering behavior between LPCVD and PECVD Si$_3$N$_4$ films using SIMS.
We deposited 200 nm-thick Si$_3$N$_4$ films via both LPCVD and PECVD techniques on Si wafers, followed by annealing at 800$^{\circ}$C for varying durations (3–40 hours).
SIMS depth profiling revealed differences in Cu gettering efficiency between the two film types.
As shown in Supplementary Figure \ref{Fig:SI_LPCVD}, PECVD-grown films accumulated approximately 10 times higher Cu concentrations compared to LPCVD films, even with shorter annealing times.
This enhanced gettering capacity arises from PECVD's characteristic high defect density and dangling bonds, generated during its low-temperature deposition process, which provide abundant Cu trapping sites.
In contrast, LPCVD films show minimal Cu concentration variation with annealing time. Their stable behavior reflects the stoichiometric perfection and high-temperature growth conditions of LPCVD, which limit available gettering sites and lead to rapid trapping saturation.

\section{Diverse gettering approaches}

\begin{figure*}[h!]
		\centering
		\includegraphics[width=\textwidth]{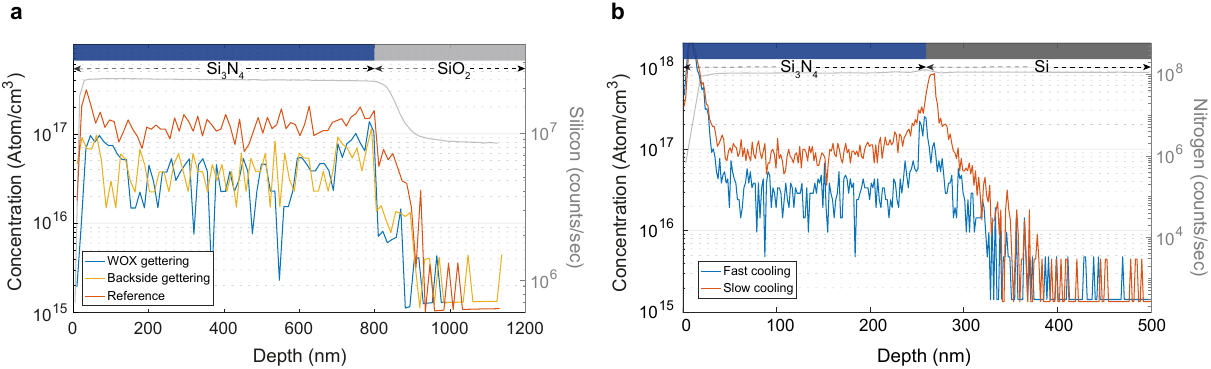}
		\caption{
		\footnotesize
		\textbf{Comparative analysis of copper gettering efficiency across diverse methods.}
		\textbf{a,} SIMS depth profiling of Cu in Si$_3$N$_4$ microresonators subjected to different gettering techniques.
		\textbf{b,} Comparative analysis of slow-cooled ($-$4$^{\circ}$C/min) and fast-cooled ($\sim -$240$^{\circ}$C/min) Si substrates coated with 260 nm PECVD Si$_3$N$_4$, after 800$^{\circ}$C 10h annealing.}
		\label{Fig:SI_get_methods}
	\end{figure*}

In addition to the gettering methods discussed in the main text, alternative techniques, such as WOX gettering and backside Si$_3$N$_4$ gettering have been explored.
Due to Cu's high diffusion coefficient $D$ in both Si and SiO$_2$ at elevated temperatures \cite{kim2006extraction}, a Si$_3$N$_4$ gettering layer can be deposited directly on Si wafers with wet-oxidized SiO$_2$ (WOX) to maintain surface smoothness.
This involves depositing a 200 nm LPCVD Si$_3$N$_4$ film on a Si wafer with 4 $\mu$m WOX, followed by annealing at 1200$^{\circ}$C for 33 hours to drive Cu impurities from the Si substrate into the Si$_3$N$_4$ waveguide. The contaminated layer is subsequently removed via dry-etching and CMP.

Backside gettering leverages residual Si$_3$N$_4$ film on the wafer backside as a gettering layer.
During Si$_3$N$_4$ device fabrication, the significant tensile stress inherent to LPCVD Si$_3$N$_4$ films often necessitates wafer bow control to mitigate stress imbalance between the wafer surfaces after front-side Si$_3$N$_4$ waveguide etching.
A thin backside Si$_3$N$_4$ layer is typically retained for stress balancing, with its thickness optimized to prevent cracking during high-temperature annealing.
This residual layer can also act as a gettering sink for Cu impurities during annealing, offering the advantage of process integration without additional steps.

Supplementary Fig. \ref{Fig:SI_get_methods}a compares Cu concentrations in Si$_3$N$_4$ microresonators across gettering methods, with Si$_3$N$_4$/SiO$_2$ fill ratio corrections applied.
Both WOX and backside gettering reduced Cu levels from $>$10$^{17}$ atoms/cm$^3$ (ppba=1380) to $\sim$4$\times$10$^{16}$ atoms/cm$^3$ (ppba=552), compared to a non-gettered reference.
While less effective than direct gettering, these methods require fewer operational steps and still achieve a notable reduction in Cu impurities in Si$_3$N$_4$ waveguides.

Another critical factor in gettering efficiency is the temperature ramp during the process.
The study in \cite{inglese2018cu} highlights the critical role of the cooling ramp during the gettering process for Cu impurities in Si using phosphorus-doped emitters.
It demonstrates that after gettering, slow cooling (e.g., $-$4$^{\circ}$C/min) between 800$^{\circ}$C and 600$^{\circ}$C significantly enhances Cu gettering efficiency, nearly eliminating Cu-related light-induced degradation.
In contrast, fast air-cooling results in weak impurity segregation, leaving residual Cu in the bulk and failing to suppress Cu-LID.
This is attributed to the enhanced Cu solubility gradient between the emitter and bulk during slow cooling, which drives efficient impurity segregation.
Thus, controlled cooling rates are essential for optimizing Cu gettering and minimizing its detrimental effects on device performance.

We investigate Cu accumulation in 260 nm PECVD Si$_3$N$_4$ films following gettering from Si substrates under two cooling regimes: controlled cooling ($-$4$^{\circ}$C/min) and rapid air cooling ($\sim -$240$^{\circ}$C/min) post-annealing.
SIMS depth profiling in Supplementary Fig. \ref{Fig:SI_get_methods}b shows a 3$\times$ higher Cu concentration in gradually cooled samples, demonstrating cooling-rate-controlled gettering efficiency.

\section{Characterization of purified Si wafers post-Cu gettering treatment}

\begin{figure*}[h!]
		\centering
		\includegraphics[width=0.8\textwidth]{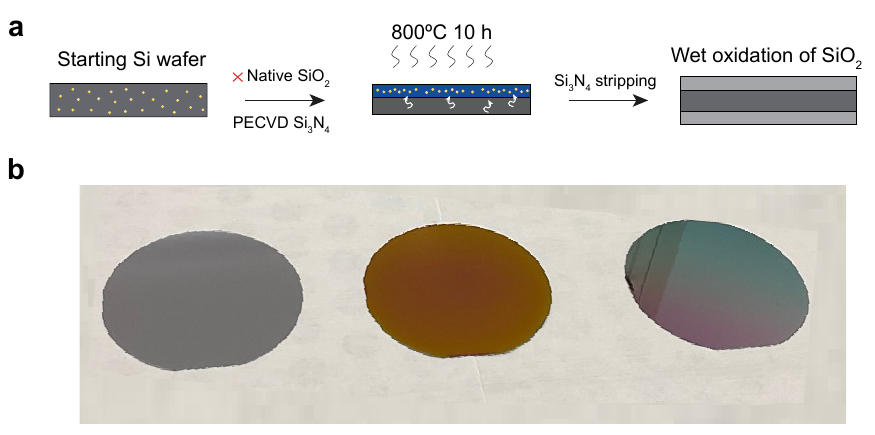}
		\caption{
			\footnotesize
			\textbf{Characterization of Si wafers after direct gettering treatment.} 
			\textbf{a,} Schematic of the direct gettering process: native SiO$_2$ stripping, front-side PECVD Si$_3$N$_4$ deposition, Si$_3$N$_4$ removal, and wet oxidation of SiO$_2$.
			\textbf{b,} Images of wafers at each stage corresponding to \textbf{a}. 
		}
		\label{Fig:SI_direct_gettering}
	\end{figure*}

	\begin{figure*}[h!]
		\centering
		\includegraphics[width=\textwidth]{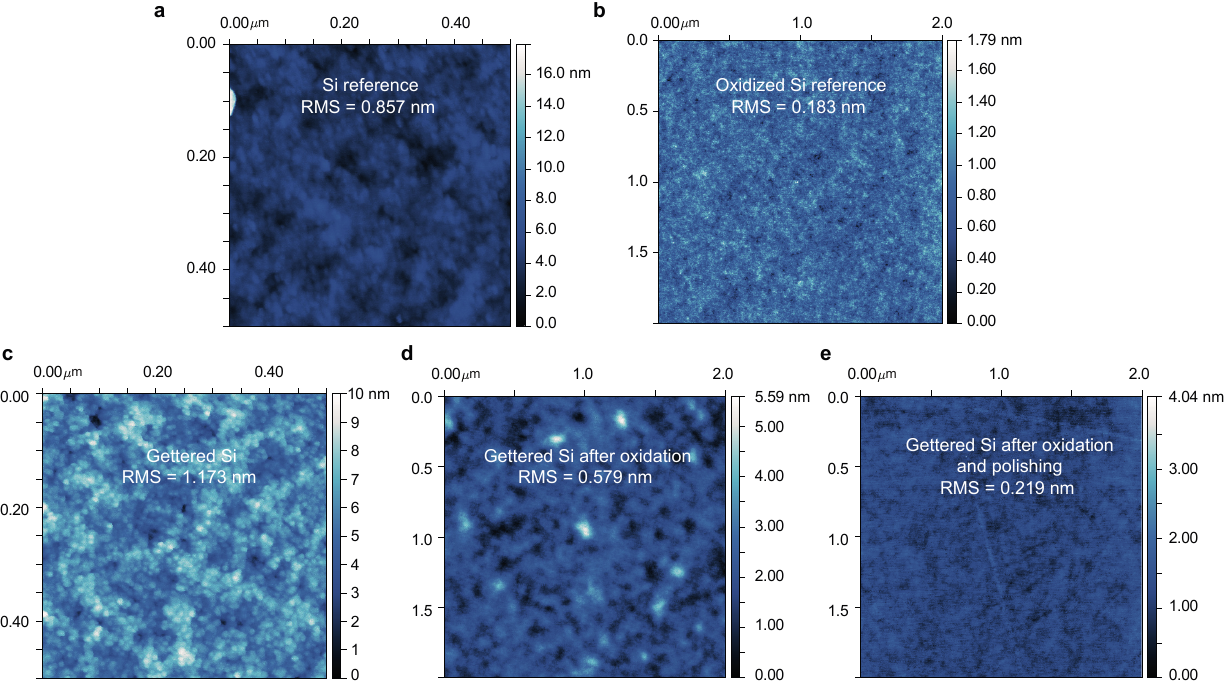}
		\caption{
			\footnotesize
			\textbf{AFM characterization of Si wafers at various gettering stages compared to a reference.} 
			\textbf{a,} Reference Si wafer.  
			\textbf{b,} Reference Si wafer after oxidation. 
			\textbf{c,} Si wafer after direct gettering.  
			\textbf{d,} Si wafer after direct gettering and oxidation.
			\textbf{e,} Si wafer after direct gettering, oxidation, and polishing. 
		}
		\label{Fig:SI_gettering_AFM}
	\end{figure*}
	
This section characterizes the surface roughness of Si wafers purified via the direct gettering approach.

Supplementary Fig. \ref{Fig:SI_direct_gettering}a illustrates the direct gettering process: native SiO$_2$ stripping with HF, front-side PECVD Si$_3$N$_4$ deposition, Si$_3$N$_4$ removal by dry- and wet-etching (HF), and wet oxidation of SiO$_2$.
Supplementary Fig. \ref{Fig:SI_direct_gettering}b provides photographs of wafers at each stage.
	
We used atomic force microscopy (AFM) to monitor surface roughness throughout this process.
After direct gettering, the root-mean-square (RMS) roughness of the Si wafer surface is 1.173 nm, reducing to 0.579 nm following the oxidation of a 4 $\mu$m SiO$_2$ layer (Supplementary Fig. \ref{Fig:SI_gettering_AFM}c, d).
	
To further minimize device loss, which correlates with SiO$_2$ surface roughness, chemical-mechanical polishing (CMP) can be applied, achieving an RMS of 0.219 nm (Supplementary Fig. \ref{Fig:SI_gettering_AFM}e). 
For comparison, a reference Si wafer from the same batch exhibits an RMS of 0.857 nm, improving to 0.183 nm after oxidation (Supplementary Fig. \ref{Fig:SI_gettering_AFM}a, b), indicating slightly better roughness than achieved with the direct gettering approach.

\begin{figure*}[h!]
		\centering
		\includegraphics[width=\textwidth]{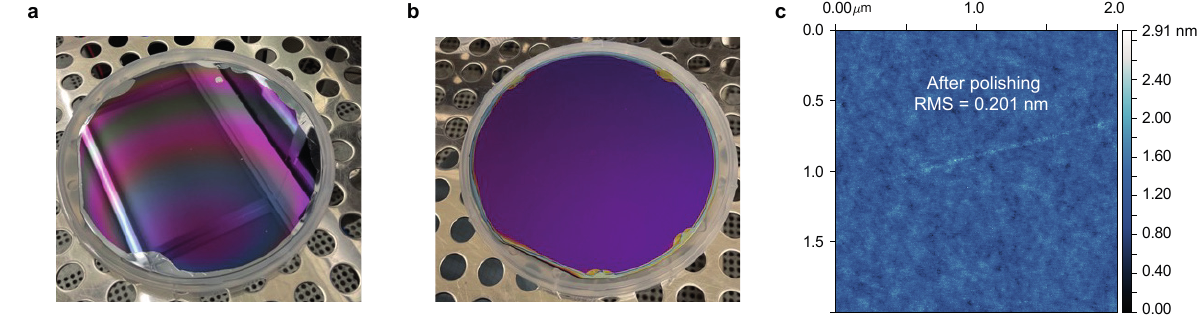}
		\caption{
			\footnotesize
			\textbf{Characterization of a Si wafer treated with the diffusion barrier approach.} 
			\textbf{a,} Front side of the Si wafer after diffusion barrier deposition (200 nm LPCVD Si$_3$N$_4$), wafer bonding, and substrate removal.
			 \textbf{b,} Backside of the wafer.
			 \textbf{c,} AFM scan of the polished top SiO$_2$ surface.
		}
		\label{Fig:SI_barrier}
	\end{figure*}
	
We characterize Si wafers treated with a diffusion barrier approach to prevent Cu diffusion into Si$_3$N$_4$ devices.
The process involves depositing 200 nm LPCVD Si$_3$N$_4$ followed by a 4 $\mu$m SiO$_2$ layer formed via wafer bonding.
Supplementary Fig. \ref{Fig:SI_barrier}a, b shows images of the wafer’s front and back sides after Si$_3$N$_4$ deposition.
AFM scan in Supplementary Fig. \ref{Fig:SI_barrier}c reveals an RMS roughness of 0.201 nm, demonstrating high-quality SiO$_2$ cladding suitable for device fabrication.

\section{Characterization of gettered Si$_3$N$_4$ PICs}

\begin{figure*}[h!]
		\centering
		\includegraphics[width=0.9\textwidth]{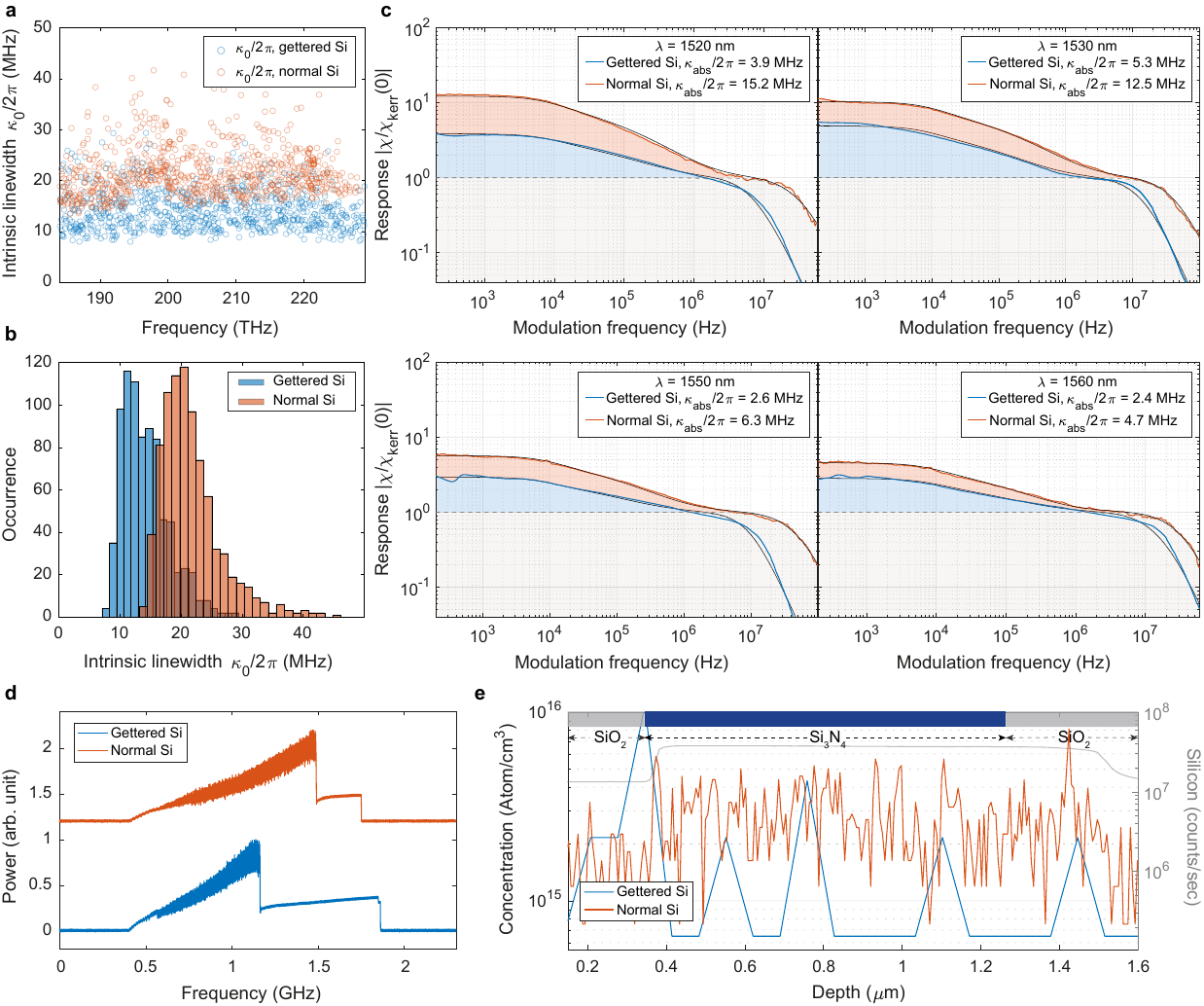}
		\caption{
		\footnotesize
		\textbf{Optical loss, thermal absorption, and SIMS characterization of gettered versus non-gettered Si$_3$N$_4$ microresonators.}
		\textbf{a,} Resonance intrinsic linewidth ($\kappa_0/2\pi$) measurements from 1310-1630 nm (TE polarization), comparing gettered (blue) and non-gettered (red, labeled "normal") devices.
		 Sample IDs: gettered (\texttt{D163\_03\_F1\_C7\_1\_04}), non-gettered (\texttt{D163\_11\_F1\_C7\_2\_04}).
		\textbf{b,} Histogram of $\kappa_0/2\pi$ for devices in \textbf{a}.
		\textbf{c,} Kerr-normalized thermal response measurements at resonances near 1520, 1530, 1550, and 1560 nm for both devices.
Colored regions denote thermal-dominated regimes at low modulation frequencies.
		\textbf{d,} Soliton step formation during pump laser scanning.
		\textbf{e,} Comparative SIMS depth profiles of Cu contamination in gettered versus non-gettered devices.}
		\label{Fig:SI_thermal}
	\end{figure*}
	
To demonstrate the effectiveness of our gettering approach in reducing optical losses in Si$_3$N$_4$ microresonators, we fabricated devices using two Si wafers from the same batch: one purified via direct gettering and the other non-gettered, ensuring identical initial bulk Cu contamination levels. Both wafers were processed identically during fabrication. The Si$_3$N$_4$ microresonators under study feature a cross-section of 2.2$\times$0.9 $\mu$m$^2$ and an FSR of 50 GHz. 

We characterized the intrinsic resonance linewidth ($\kappa_0/2\pi$) of both devices across the wavelength range of 1310 nm to 1630 nm (Supplementary Fig. \ref{Fig:SI_thermal}a, b). The gettered device exhibited a mean $\kappa_0/2\pi$ of 14.2 MHz and a most probable value of 11 MHz, compared to 21.6 MHz and 20 MHz for the non-gettered device, representing a $\sim$7-9 MHz reduction in intrinsic linewidth due to gettering.

To further evaluate thermal effects, we employed Kerr-calibrated resonance frequency response measurements.
Supplementary Fig. \ref{Fig:SI_thermal}c illustrates the measured responses and fittings for thermal (blue and red) and Kerr (grey) contributions.
The thermal-to-Kerr ratio, $\gamma = \chi_{\mathrm{therm}}(0)/\chi_{\mathrm{Kerr}}(0)$, demonstrates significantly lower thermal absorption in gettered device, with $\gamma$ values 3-to-5-fold higher in non-gettered device, highlighting the efficacy of the gettering process in minimizing thermal losses in Si$_3$N$_4$ microresonators.
Residual thermal absorption, attributed to hydrogen-related bonds (e.g., Si-H at 1517 nm and N-H at 1530 nm) \cite{pfeiffer2018ultra}, is evident in the increased $\gamma$ values near 1520 nm and 1530 nm. Such residual absorption could, in principle, be further reduced through more stringent thermal treatments, such as higher-temperature annealing.

While our Si$_3$N$_4$ devices are currently limited by scattering losses, the gettering treatment, which significantly reduces thermal absorption and facilitates soliton generation, plays a critical role in applications where scattering losses are already optimized to near-fundamental limits. This underscores the broader applicability of our approach in advancing ultra-low-loss integrated photonic systems.

Supplementary Fig. \ref{Fig:SI_thermal}d reveals a significantly extended soliton step in the gettered device (705 MHz) compared to the non-gettered device (256 MHz) during pump laser scanning.
For gettered Si$_3$N$_4$ samples, we present raw SIMS measurements without applying the Si$_3$N$_4$/SiO$_2$ fill ratio correction in Supplementary Fig. \ref{Fig:SI_thermal}e. Since no Cu signal is detected in the gettered sample, corrections would unnecessarily complicate the analysis and risk misrepresenting the absence of impurities.
In contrast, non-gettered samples show measurable Cu contamination (5$\times$10$^{15}$ atoms/cm$^3$).
These results demonstrate the benefits of gettering: effective impurity suppression and consequent improvements in optical performance (reduced loss, suppressed thermal effects, and extended soliton steps).

\section{Cu impurities in commercial Si$_3$N$_4$ PICs}

\begin{figure*}[h!]
		\centering
		\includegraphics[width=\textwidth]{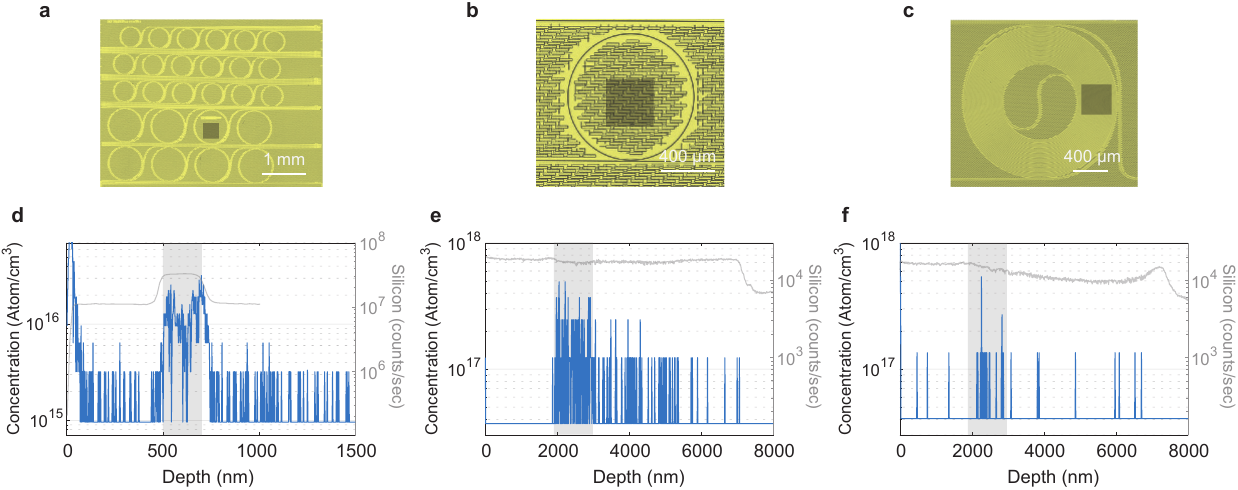}
		\caption{
		\footnotesize
		\textbf{SIMS characterization of Cu concentration in commercial Si$_3$N$_4$ samples (Ligentec SA).}
		\textbf{a-c,} Microscopic images of Si$_3$N$_4$ devices under test, with the grey square indicating the measurement spot. All samples were fabricated in a commercial foundry (X-FAB).
		\textbf{d-f,} The corresponding depth profiles of Cu concentration, with the Si$_3$N$_4$ layer highlighted in grey. The Si$_3$N$_4$ layer is identified using silicon secondary ion intensity and layer structure data provided by Ligentec SA.}
		\label{Fig:SI_LGT}
	\end{figure*}

We examined the Cu concentration in commercial Si$_3$N$_4$ samples (Ligentec SA) fabricated in a contamination-controlled commercial foundry (X-FAB) to confirm the pervasive presence of Cu in Si$_3$N$_4$ photonic devices.
Despite stringent contamination protocols, SIMS depth profiles in Supplementary Fig.\ref{Fig:SI_LGT}d-f reveal Cu impurities at 10$^{16}$-10$^{17}$ atoms/cm$^3$ level within the Si$_3$N$_4$ layer.
This suggests that Cu contamination is not solely attributable to external fabrication processes but is fundamentally tied to the bulk Si wafers used as substrates.

\section{Cu impurities in LNOI/LTOI platforms}

\begin{figure*}[h!]
		\centering	\includegraphics[width=0.9\textwidth]{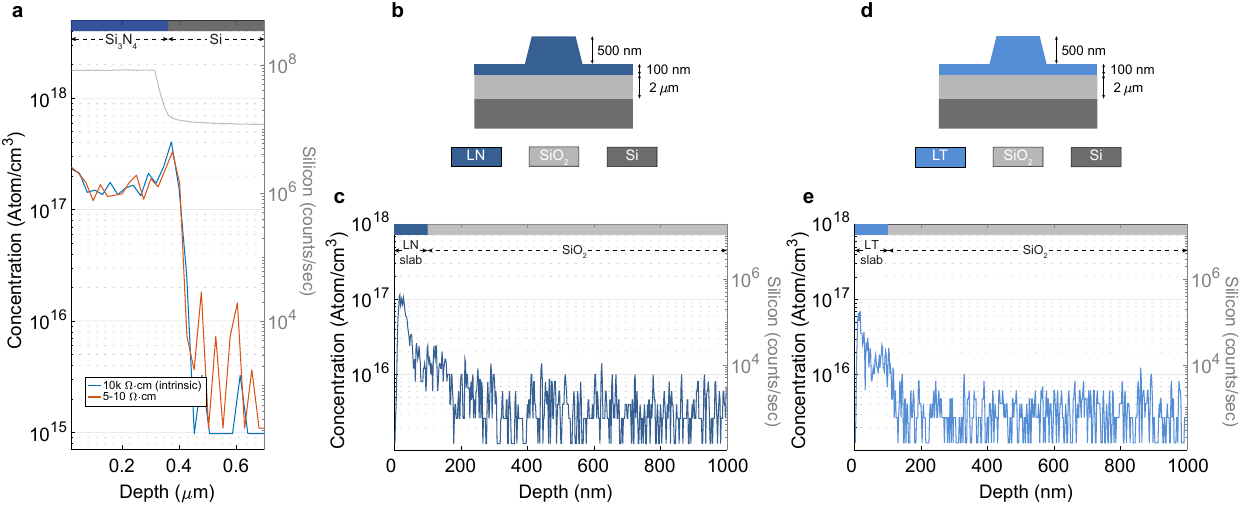}
		\caption{
		\footnotesize
		\textbf{SIMS depth profiles of Cu concentration in intrinsic, B/P-doped Si wafers, and LNOI/LTOI platforms.}
		\textbf{a,} SIMS depth profiles of Cu in intrinsic and B/P-doped Si wafers.
		\textbf{b,} Schematic of the tested LNOI waveguide (500 nm core, 100 nm slab).
		\textbf{c,} SIMS depth profile of Cu in LNOI chip, showing presence of Cu in the 100 nm slab (waveguide omitted due to low density).
		\textbf{d,} Schematic of the tested LTOI waveguide.
		\textbf{e,} SIMS depth profile of Cu in LTOI, with detection similarly confined to the 100 nm slab.
		}
		\label{Fig:SI_LN_LT}
	\end{figure*}
	
To extend our investigation of Cu impurities to integrated photonic platforms, we performed SIMS depth profiling on lithium niobate-on-insulator (LNOI) and lithium tantalate-on-insulator (LTOI) waveguides. 
We first assessed Cu contamination in intrinsic Si wafers (resistivity $>$10k $\Omega\cdot$cm) commonly used in LNOI/LTOI fabrication.
A thin Si$_3$N$_4$ layer was deposited on intrinsic Si wafers and annealed at 1200$^{\circ}$C for 11 hours in N$_2$ to promote Cu gettering in the Si$_3$N$_4$ layer.
Supplementary Figure \ref{Fig:SI_LN_LT}a compares Cu concentrations in an intrinsic Si wafer and a boron-doped p-type Si wafer (5–10 $\Omega\cdot$cm) from the same supplier.
Comparable Cu levels were observed in both wafer types, indicating pervasive contamination in commercial Si, even in high-purity (7N–8N, 10–0.1 ppbw) intrinsic substrates. However, the measured Cu concentration ($>$10$^{17}$ atom/cm$^3$, $>$1050 ppba) exceeds typical bulk impurity levels, suggesting extrinsic contamination, likely introduced during Si wafer polishing.

Next, we analyzed Cu in LNOI/LTOI structures with 500 nm waveguide and 100 nm slab heights using Cs$^{+}$ sputtering. Supplementary Figure \ref{Fig:SI_LN_LT}c, e show that Cu is confined to the top 100 nm of the slab, with negligible detection in the SiO$_2$ cladding. This localization arises because the measurement region primarily samples the slab due to the sparse distribution of waveguides in patterned chips. The Cu concentration in the slab differs from that in intrinsic Si, possibly due to variations in substrate sourcing or differences in gettering efficiency between Si$_3$N$_4$ and LN/LT.

The detection of Cu in LNOI/LTOI platforms confirms the ubiquitous nature of Cu contamination in silicon-based integrated photonic systems, emphasizing the need for stringent control during Si wafer polishing, handling, and integration to minimize impurity-induced losses in photonic devices.

\clearpage
%%%%%%%%%%%%%%%%%%%%%%%%%%%%%%%%%%%%%%%%%%%%%%%%%%%%%%%%%%%%%%%%

\noindent \textbf{Acknowledgments}:
The XPS measurements were performed by Dr. Mounir Mensi.

\bigskip
\bibliographystyle{apsrev4-2}
\bibliography{SI}